\documentstyle[preprint,tighten,version2,aps]{revtex}
\def\goto{\mathop{\;\longrightarrow\;}}
\def\princint{\ \raise 0.55 ex\hbox to 0.7em{\hrulefill}\kern-1em\int}
{\makeatletter%
\gdef\labeleqs#1{{%
\edef\@currentlabel{%
\ifappendixon\appletter\fi
\ifsecnumbers\ifnum\c@secnum>0
\arabic{secnum}.\fi\fi\arabic{equation}}%
\label{#1}%
}}%
}
\begin{document}
\draft
\preprint{BROWN-HET- 1005}
\smallskip
\preprint{BU-CCS-950602} 
\smallskip
\preprint{IFUP-TH 8/95}

\begin{title}
Critical Behavior of Simplicial Chiral Models
\end{title}
\author{Richard C. Brower$^1$, Massimo Campostrini$^2$, Kostas Orginos$^3$, \\
Paolo Rossi$^2$, Chung-I Tan$^3$, and Ettore Vicari$^2$}
\begin{instit}
$^1$ Physics Department, Boston University, Boston, MA 02215, USA.\\
$^2$ Dipartimento di Fisica dell'Universit\`a and
I.N.F.N., I-56126 Pisa, Italy.\\
$^3$ Department of Physics, Brown University, Providence RI 02912, USA.
\end{instit}
\begin{abstract}
The large-$N$ saddle-point equations for the principal chiral models
defined on a $d-1$ dimensional simplex are derived from the external
field problem for unitary integrals.  The saddle point equation are
studied analytically and numerically in many relevant instances,
including $d=4$ and $d\rightarrow\infty$, with special attention to
the critical domain, which is found to correspond to $\beta_c=1/d$ for
all $d$.  Related models (chiral chains) are discussed and large-$N$
solutions are analyzed. 

\end{abstract}

% ========================= BODY =========================

\narrowtext

\section{Introduction}
\label{intro} 

The $1/N$ expansion of matrix-valued field theories is probably the
most important non-perturbative and non-numerical theoretical tool
presently available in the study of such models as non-Abelian gauge
theories and two-dimensional quantum gravity.  A resolution of the
above-mentioned models in the large-$N$ limit would be the starting
point for many analytical developments. In particular when a lattice
formulation is involved one must consider different possibilities in
the search for the continuum limit; for the case of asymptotically
free theories one must explore the limit of vanishing coupling $g=0$
(trivial fixed point) while keeping a physical mass scale fixed, while
in the case of quantum gravity one must search for a nontrivial fixed
point $g_c$ and reach the limit with a specific power-law dependence
on $N$ of $g-g_c$, which is known as ``double scaling
limit''~\cite{Douglas,GrossM,BrezinK}.  Therefore it is useful to
achieve a full knowledge of the coupling dependence of such models,
from extreme weak coupling to strong coupling, in order to explore
those regions that may turn out to be physically most interesting.

As a matter of fact, notwithstanding many recent efforts toward an
understanding of the possible properties of large-$N$ solutions to
nontrivial quantum field theories, our present analytical knowledge is
limited to a small number of few-matrix systems.  This number is even
smaller if we restrict our attention to the case of unitary matrix
fields, which is especially relevant to the problem of lattice QCD.
To the best of our knowledge, the only solved examples are
Gross-Witten's single-link problem~\cite{Gross} and its generalizations, the
external field problem~\cite{Brower,BrezinG} and $L=3,4$ chiral
chains~\cite{BRT,Friedan}.

We stress that extending the number of solved few-matrix systems is
not at all a pointless exercise. Indeed apart from purely theoretical
informations that might be achieved, not  only does every few-matrix system
have a reinterpretation, via the double scaling limit, as some
different kind of matter coupled to 2-dimensional quantum gravity, but
also every few-matrix system involving unitary matrices can be
reinterpreted as the generating functional for a class of integrals
over unitary groups, and these integrals in turn are the essential
missing ingredient in the context of a complete algorithmization of
the strong coupling expansion of many interesting models~\cite{SC}.

Following Ref.~\cite{SC}, we may introduce the notion of a
``superskeleton'', that is a graph whose vertices are joined by at
most one link (simple graph).  As has been shown, knowledge of all
the group integrals involved in the strong coupling expansion of a
lattice model with nearest-neighbor interactions defined on such a
graph provides sufficient information for the algorithmic
reconstruction of the strong coupling series for a model enjoying the
same global symmetry and defined on an arbitrary lattice.

These were basic motivations for us to begin the
study of the class of lattice chiral models which we termed
``simplicial chiral models''~\cite{RossiTan}.  In particular we
focused on principal chiral models, with a global $U(N)\times U(N)$
symmetry, defined on a $d-1$ dimensional simplex formed by connecting
$d$ vertices by ${(d-1)(d-2)/2}$ links, and explored specifically
the large-$N$ limit of such models, whose relevance we have just been
discussing.  

Our fundamental result is the reduction of the above
problem to that of solving a single inhomogeneous integral equation
for the eigenvalue distribution of a single Hermitian semi-positive
definite matrix. Although we could not find a closed form solution to
this equation for arbitrary $d$, we are able to solve it in several
interesting special cases and we set up a systematic numerical
approach to the solutions which led us to a conjecture about the
location of the critical surface as a function of $d$. We have also studied
in detail the related topic of chiral chains, their strong coupling
expansion and critical behavior. As a  result of these analyses,
we are confident that the critical surface is  defined by
$\beta_c = 1/d$ for all $d$. 

Moreover, by treating $d$ as a continuous parameter, there are two
distinct regions. For small $d$, $0< \lowercase{d} < 4$, the models
exhibit the third order Gross-Witten transition. Indeed for $d =
1,2,3$ they coincide exactly with the chiral chains studied earlier by
Brower, Rossi and Tan~\cite{BRT}.  In this region, the criticality is
related to that of $O(n)$ spin models on random surfaces, as discussed
by Gaudin-Kostov~\cite{Gaudin}. For $\lowercase{d} > 4$, however,
there is a first order transition ending at the ``upper critical''
dimensions $\lowercase{d}= 4$, which we scrutinize in some detail.

This paper is organized as follows:
\begin{itemize}
\item In Section~\ref{matrix} we set up our formalism for simplicial chiral
model and derive the large-$N$ effective action and a representation
for the internal energy.  We begin in Section~\ref{spin} by illustrating
the formalism for the simpler case of vector $O(N)$ spin models on a
simplicial lattice, deriving the closed form large-$N$ solution for
arbitrary values of $d$ and studying its properties.
Sec~\ref{largeNlimit} gives the large-$N$ effective action for the
simplicial chiral model and Section~\ref{saddlepoint}, the saddle-point
equation (large-$N$ Schwinger-Dyson equation) for the eigenvalue
distribution, discussing its features and converting it into a
standard inhomogeneous Fredholm equation of the second kind.

\item In Section~\ref{sec5} we analyze the solvable examples. We begin with
integer values of $d< 4$ (which correspond to $L<4$ chiral chains),
followed in Section~\ref{sec6} by a detailed discussion of the $d=4$
case, including features of the weak and strong coupling expansions
and the asymptotic expansion around the critical point $\beta_c=1/4$.
We present in Section~\ref{sec8} the exact result for the limit
$d\rightarrow\infty$ and develop in Section~\ref{sec8a} a treatment
based on the $1/d$ expansion. This large-$d$ analysis, which works
best outside the critical region: $\beta \gg \beta_c$, has provided us
with the first numerical confirmation for the conjecture that
$\beta_c=1/d$, and is discussed in greater details in Appendix D.

\item In Section~\ref{sec7} we solve the models at criticality for arbitrary
$0< d\leq 4$ and sketch the peculiar features of the critical behavior
when $d>4$. Also in Section~\ref{sec9a} we present numerical methods
based on the Gaussian integration techniques for $d>4$ on the critical
surface $\beta_c = 1/d$.

\item Some technical extensions are included in other appendices.
Appendix~\ref{appa} is devoted to a discussion of the double-scaling
limit of critical chiral chains $L\leq 4$ and
Appendix~\ref{appb} extends the discussion of chiral chains to $L\geq
4$ by an analysis of strong coupling expansion.  Whereas
Appendix~\ref{appc} is devoted to details of the weak and strong
coupling expansions and series analysis for the $d=4$ simplicial
chiral model.
\end{itemize}

\section{Simplicial Chiral Models}
\label{matrix}

A $d-1$ dimensional simplex is formed by connecting in a fully
symmetric way $d$ vertices by ${(d-1)(d-2)/ 2}$ links. Let us assign a
$U(N)$ matrix to each vertex. The partition function for principal
chiral models on a simplicial lattice is obtained by integrating over
unitary $N\times N$ matrices with a normalized invariant Haar measure:
\begin{equation}
Z_d\;=\; \int \prod_{i=1}^d d U_i\exp \left\{ N\beta \sum_{i>j=1}^d
{\rm Tr} \, \left[ U_i U^\dagger_j + U_j U_i^\dagger\right]\right\}\;.
\label{zdm}
\end{equation} 
Thus the $d$-matrix simplicial model has an underlying permutation
symmetry instead of the cyclic symmetry of the $d$-matrix chiral chains.
For $d$ = 1, 2 and 3 these two symmetries and the associated
models are equivalent.  We shall explore the $U(N)\times U(N)$
symmetry of the system and, in particular, study its critical behavior
in the large-$N$ limit.

For this purpose, it is sufficient to study the bulk ``thermodynamic"
properties, {\it e.g.}, the free energy density, internal energy, and specific
heat, which  are respectively given by
\begin{eqnarray}
&&F_N(\beta,d) \;=\; {1\over N^2}\ln Z_d\;,\nonumber \\
&&U_N(\beta,d)\;=\; {1\over 2}{\partial F_N\over \partial\beta}\;,\nonumber \\
&&C_N(\beta,d)\;=\; \beta^2{\partial U_N\over \partial \beta}\;.
\label{ae2}
\end{eqnarray}
We  focus in this paper on computing the free energy and determine
the critical point $\beta_c$ for all values of the parameter $d$ in the
large $N$-limit. We find that there is a sequence of critical theories
with $\beta_c = 1/d$, which exhibit a third order Gross-Witten
singularities for $d < 4$ and a first order transition for $d >
4$. Special attention will be give to the marginal dimension at
$d=4$. Scaling exponents and finite size (or double scaling)
properties will be presented in some special cases, but a thorough
investigation for all $d$ is beyond the scope of this paper.

\subsection{External Field Method}
\label{extFM}

Because of the the permutation symmetry of the vertices, the
simplicial chiral models can be reformulated in terms of a Lagrange
multiplier field which decouples the original degrees of freedom.  The
resulting effective theory is very reminiscent of the mean field
approximation to standard lattice models, but in contrast with mean
field this reformulation is exact.

We therefore  replace the direct interaction between unitary matrices
with the coupling to an auxiliary field, which in this case is a complex
$N\times N$ matrix, $A$, introduced in the following representation
of the identity,
\begin{equation}
1\;=\;{
\int d A \exp\left\{ -N\beta\,{\rm Tr} \,\left[ \left( A - \sum_{i=1}^d U_i\right)
\left( A^\dagger - \sum_{i=1}^d U_i^\dagger\right)\right]\right\}
\over
\int d A \exp \left\{-N\beta\,{\rm Tr} \,\left[ A 
A^\dagger \right]\right\}}\;.
\label{unv}
\end{equation}
Again by exchanging the order of integrations and representing the partition function
in the form $Z_d=\widetilde{Z}_d/\widetilde{Z}_0$, we may obtain 
\begin{equation}
\widetilde{Z}_d\;=\;\int d A
\int \prod_{i=1}^d dU_i 
\exp \left\{ N\beta {\rm Tr} \left[
 -\left[AA^\dagger\right] +  \left[A\sum_i U_i^\dagger\right]
+ \left[A^\dagger \sum_i U_i\right] - d\right]\right\}\;.
\label{tzm}
\end{equation}
By performing the single-link external field integral, we may introduce
the auxiliary function,
\begin{equation}
F_N(BB^\dagger)\;=\; {1\over N^2} \ln
\int d U \exp \left\{{N\over 2} {\rm Tr} \left[ BU^\dagger +
UB^\dagger\right]\right\} \; ,
\label{fnv}
\end{equation}
and re-express $\widetilde{Z}_d$, up to an irrelevant multiplicative 
factor, in the form
\begin{equation}
\widetilde{Z}_d\;=\; \int dB\exp \left[
-{N\over 4\beta}{\rm Tr} BB^\dagger + N^2dF(BB^\dagger)-N^2\beta d\right]\;,
\label{tzm2}
\end{equation}
where $B$ replaces $2\beta A$.

A first crucial point in our analysis is the observation that the
integrand in Eq.~(\ref{tzm2}) is a function of the eigenvalues $x_i$
of the Hermitian semi-positive definite matrix $BB^\dagger$. Moreover
Morris~\cite{Morris} has shown that, when integrating over complex
matrices, a proper parameterization may offer in specific cases, like
ours, the possibility of performing the ``angular'' integrations
exactly and reducing the problem to that of integrating over the $N$
variables $x_i$. Referring to Morris' paper for a proof of the angular
integration, we apply it to our Eq.~(\ref{tzm2}) thus obtaining, again
up to irrelevant numerical factors,
\begin{equation}
\widetilde{Z}_d\;=\; 
\int_0^\infty \prod_{i=1}^d d x_i \prod_{j> i}(x_i-x_j)^2
\exp \left[ -{N\over 4\beta}\sum_i x_i + N^2 d F(x_i) - N^2 \beta d\right]\;.
\label{tzm3}
\end{equation}
The second crucial observation concerns the function $F(x_i)$. It
is known exactly for all $U(N)$ groups~\cite{BRT}, while integral
representations exist for $SU(N)$ groups~\cite{BRT2}, and it takes on a
relatively simple form in the large-$N$ limit.   Before proceeding further, we
shall first provide an even simpler illustrative example whose large-$N$ solution
can be obtained fairly straightforwardly.

\subsection{Pedagogical example: simplicial spin models.}
\label{spin}

Consider instead of our simplicial chiral models, an example of an
$O(N)$ symmetric nonlinear model defined on a simplex. The same basic
methods used for the chiral models are easily illustrated in this much
simpler context. 

The partition function is obtained by integrating
over the $N-1$ independent components of $d$ vectors,
\begin{equation}
Z_d\;=\; \int \prod_{i=1}^d d {\bf s}_i 
\;\delta ({\bf s}_i^2 -1) \exp \left[ N\beta \sum_{i>j=1}^d 
{\bf s}_i \cdot {\bf s}_j\right]\;.
\label{zds}
\end{equation}
The effective field is a single unconstrained $N$-component vector,
${\vec z}$, which again can be introduced as Lagrange multiplier field
via an identity,
\begin{equation}
1\;\equiv\;{
\int d {\bf z} \exp - {N\beta\over 2} ({\bf z}-\sum_{i=1}^d {\bf s}_i)^2
\over 
\int d {\bf z} \exp - {N\beta\over 2} {\bf z}^2}\;.
\label{1e1}
\end{equation}
Upon substituting this identity into Eq.~(\ref{zds}) and
inverting the order of the integrations, we may then represent the
partition function by $Z_d=\widetilde{Z}_d/\widetilde{Z}_0$, where
\begin{equation}
\widetilde{Z}_d\;=\; \int d{\bf v} \prod_{i=1}^d \int d {\bf s}_i
\;\delta ({\bf s}_i^2 -1) 
\exp \left[ - {N\beta\over 2}{\bf v}^2 + N\beta\, 
{\bf v}\cdot \left( \sum_{i=1}^d {\bf s}_i\right) - {N\beta\over 2}d\right]\;.
\label{ztds}
\end{equation}
It is now possible to perform the decoupled constrained integrations.
To this end we may define the auxiliary function
\begin{equation}
F_N(z^2)\;=\;
{1\over N} \ln \int d {\bf s} \;\delta ({\bf s}^2 -1) 
\exp N {\bf s}\cdot {\bf z} \;,
\label{fns}
\end{equation}
where $z^2={\bf z}\cdot{\bf z}$.  This function is known explicitly
for all values of $N$ and admits a large $N$ limit~\cite{Stanley}:
\begin{eqnarray}
F_N(z^2)&&=\; {1\over N} \ln \;\Gamma\left({N\over 2}\right) \;I_{{N\over 2}-1}(Nz)
\left( Nz\over 2\right)^{1-{N\over 2}} \nonumber \\
&&\goto_{N\rightarrow\infty}
{1\over 2} \left[ \sqrt{1+4z^2} - 1  - \ln\left({1\over 2}\sqrt{1+4z^2}+{1\over 2}\right)
\right]\;.
\label{fns2}
\end{eqnarray}
As a consequence for large $N$ we have the following representation of
$\widetilde{Z}_d$,
\begin{equation}
\widetilde{Z}_d\;=\; \int d {\bf z}
\exp \left\{-{N\over 2} \left[
{z^2\over \beta} - d\left( \sqrt{1+4z^2} -1 - \ln\left( {1\over 2}
\sqrt{1+4z^2}+{1\over 2}\right)\right) + \beta d \right]\right\}\; .
\label{ztds2}
\end{equation}
The large-$N$ value of the integral in Eq.~(\ref{ztds2}) may be obtained
by a saddle-point estimate. After some simple manipulations, the
saddle-point equation may be reduced to
\begin{equation}
1 - {z^2\over \beta} \;=\; {d\over 2}\left( 1-\sqrt{1+4z^2}\right)\;.
\label{spe}
\end{equation}
The solution of this equation is 
\begin{equation}
z^2\;=\; \beta\left[ 1-{d\over 2}+ \beta {d^2\over 2}
+{d\over 2}\sqrt{ \left( 1-\beta d\right)^2 + 4\beta } \,\right]\;.
\label{sps}
\end{equation}
By taking the logarithmic derivative of the partition function with respect
to $\beta$, we obtain an expression for the internal energy (per unit link) $U_d$ 
of simplicial spin models in the large-$N$ limit
\begin{equation}
{d(d-1)\over 2} U_d\;=\;
{1\over 2}\left( {z^2\over \beta^2}-d-{1\over \beta}\right)\;=\;
{d\over 2}\left[ {\beta d -1\over 2\beta}+
{1\over 2\beta}\sqrt{ (1-\beta d)^2 + 4\beta} -1\right]\;.
\label{energys}
\end{equation}
We may check many special cases of this result, and in particular we
may notice that the r.h.s. of Eq.~(\ref{energys}) is zero when $d=1$,
while when $d=2$
\begin{equation}
U_2\;=\; {1\over 2\beta} \left[ \sqrt{1+4\beta^2} -1\right]\; ,
\label{enes2}
\end{equation}
consistent with the single-link model result.

Finally let us notice that in the large-$d$ limit, as a trivial
consequence of the structure of the model, the solution we found
coincides with the mean field solution, which is exact in this limit.
It is worth observing that, while Eq.~(\ref{energys}) is formally
correct for all values of $\beta$, in order to recover the standard
strong and weak coupling expansions of the solution we must separately
consider the two different regimes $\beta<\beta_c$ and $\beta >
\beta_c$, where for all $d$ we obtain $\beta_c=1/d$.

\subsection{Large-$N$ limit}
\label{largeNlimit}

Returning to the simplicial chiral models, we are again interested in
the large-$N$ limit. For the free energy function $F(x_i)$ resulting from a
one-link integral over a $U(N)$ matrix, the limiting form can be
extracted by solving the Schwinger-Dyson equations and written in a
simple closed form~\cite{Brower,BrezinG},
\begin{equation}
F(x_i)\;=\;
{1\over N} \sum_i \sqrt{r+x_i} - {1\over 2N^2}
\sum_{ij} \ln \left( {\sqrt{r+x_i}+\sqrt{r+x_j}\over 2}\right)
-{r\over 4} - {3\over 4}\;.
\label{fxi}
\end{equation}
We must distinguish two different phases,
a weak coupling regime where $r=0$ and 
\begin{equation}
{1\over N}\sum {1\over \sqrt{x_i}}\;\leq \;1,
\label{wcond}
\end{equation}
 and a strong coupling regime where $r$ is dynamically determined by
the condition,
\begin{equation}
{1\over N} \sum_i{1\over \sqrt{r+x_i}}\;=\;1 \; .
\label{cond}
\end{equation}
It is  important for future developments to observe
that Eq.~(\ref{cond}) also leads to the condition
\begin{equation}
{\partial F(x_i,r)\over \partial r}\;=\;0\;.
\label{cond2}
\end{equation}
It is completely legitimate to apply the above results to a saddle-point evaluation
of the large-$N$ limit of the integral appearing in Eq.~(\ref{tzm3}).
To this end we may define an effective action
\begin{equation}
S_d\;=\; {N\over 4\beta}\sum_i x_i - N^2 d F(x_i) -
\sum_{i\neq j}\ln (x_i-x_j) + N^2\beta d\;,
\label{seff}
\end{equation}
and derive a saddle-point equation
\begin{equation}
{1\over N} {\partial S_d\over \partial x_i}\;=\;
{1\over 4\beta} - N d {\partial F\over \partial x_i} - 
{2\over N}\sum_{i\neq j} {1\over x_i- x_j}\;=\;0\;.
\label{sdeq}
\end{equation}
Very simple manipulations, including the use of Eq.~(\ref{cond2}),
lead to a reformulation of Eq.~(\ref{sdeq}), which can be turned
into the relationship 
\begin{equation}
{\sqrt{r+x_i}\over 2\beta} - d \;=\;
{1\over N} \sum_{i\neq j}
{(4-d)\sqrt{r+x_i} + d \sqrt{r+x_j}\over x_i-x_j}\;.
\label{sdeq2}
\end{equation}
This equation, supplemented with the condition $x_i\geq 0$ and with
the constraint $r=0$ (weak coupling) or Eq.~(\ref{cond}) (strong
coupling) is the fundamental saddle-point equation of principal chiral
models on a simplicial lattice. It is the starting point of most of
the developments presented in the following sections.

We recall that, once Eq.~(\ref{sdeq2}) is solved, knowledge of the
saddle-point value of the eigenvalues $\bar{x}_i$ allows the large-$N$
evaluation of $\widetilde{Z}_d$ via the relationship,
\begin{equation}
\widetilde{Z}_d\;\goto \exp[ - S_d(\bar{x}_i)] \; ,
\label{Zdsp}
\end{equation}
and we can also extract the internal energy per unit
link by taking a logarithmic derivative of $Z_d$ with respect to $\beta$
which leaves us with the relationship,
\begin{equation}
d(d-1)\, U\;=\; {1\over 4\beta^2} \sum_i \bar{x}_i
-d -{1\over \beta}\;.
\label{enem}
\end{equation}

\subsection{Saddle-point equation.}
\label{saddlepoint}

In order to study Eq.~(\ref{sdeq2}) we shall start by applying
well-established techniques, and in particular by introducing an eigenvalue
density function.  It is however convenient first to introduce a new variable
$z_i$ whose formal definition is
\begin{equation}
z_i\;=\; \sqrt{r+x_i}\;.
\label{s4e1}
\end{equation}
subject to the condition $0\leq \sqrt{r} \leq z_i$.  We may assume that the
eigenvalue variable $x_i$ lies in a single interval $[x_a,x_b]$, $0\leq
x_a\leq x_b$.  In terms of the new variable $z_i$, one has $z_i\epsilon [a,b]$
where $a=\sqrt{ r+x_a}$, $b=\sqrt{r+x_b}$ and
\begin{equation}
0\leq \sqrt{r} \leq a\leq b.
\label{s4e2}
\end{equation} 
For weak coupling, $r=0$, and  we expect in general $a={\sqrt x_a} >0$. 
For strong coupling, one expects $x_a=0$ so that  $a={\sqrt r}\neq 0$. 

We shall be interested in the weak-strong transition as one varies $d$
and $\beta$. In a third-order transition, typical of large-$N$
transition previously studied, $a_c={\sqrt r_c}= 0$. In a first-order
transition, which we will encounter for $d>4$, $a_c={\sqrt r_c}\neq 0$
when approached from the strong coupling regime.

Denoting the large-$N$ eigenvalue density by $\rho(z)$; it vanishes outside
the interval $[a,b]$.  We may now turn Eq.~(\ref{sdeq}) into the following
integral equation,
\begin{equation}
{z\over 2\beta} - d\;=\;\princint_a^b dz^\prime \,\rho(z^\prime)
\left( {2\over z-z^\prime} - {d-2\over z+z^\prime}\right)\;.
\label{s4e3}
\end{equation}
The function $\rho(z)$, and therefore also the extremes $a$ and $b$
of the integration region, are thus determined dynamically. In particular
the normalization condition,
\begin{equation}
\int_a^b d z^\prime\rho(z^\prime)\;=\;1\;,
\label{s4e4}
\end{equation}
must be satisfied. 

In addition to the positivity requirement, $\rho(z)\geq 0$ over the
interval $[a,b]$, the desired solution to Eq.~(\ref{s4e3}) must also
satisfy either the weak coupling inequality, Eq.~(\ref{wcond}) or the
strong coupling constraint, Eq.~(\ref{cond}). In the large-$N$ limit,
Eq.~(\ref{wcond}) becomes
\begin{equation}
\int_a^b d z^\prime {\rho(z^\prime)\over z^\prime} \;\leq \;1\;,
\label{s4e5w}
\end{equation}
whereas Eq.~(\ref{cond}) becomes 
\begin{equation}
\int_a^b d z^\prime {\rho(z^\prime)\over z^\prime} \;=\;1\;.
\label{s4e5}
\end{equation}
The determination of the transition point, $\beta_c$, and of the
critical behavior around this value is one of the interesting physical
problems concerning this model.

Eq.~(\ref{s4e3}) has a somewhat unconventional form when compared to
other integral equations, because of the special structure of its
kernel.  We may however perform a few manipulations in order to obtain
a more familiar relationship. Our starting point is the introduction
of an analytic function of $z$, by the definition
\begin{equation}
f(z)\;\equiv\; \int_a^b {\rho(z^\prime)\over z-z^\prime} d z^\prime\;.
\label{s4e6}
\end{equation}
By construction, the analyticity domain of $f(z)$ is the complex $z$
plane with the exception of a cut on the positive real axis in the
interval $[a,b]$.  The discontinuity on the cut may be parameterized
by writing
\begin{equation}
f(z\pm i\epsilon)\;=\; R(z)\mp i\pi\rho(z)\;,
\label{s4e7}
\end{equation}
when $z\in [a,b]$ and it is easy to recognize that
\begin{equation}
R(z)\;=\; {z\over 4\beta} - {d\over 2} + {d-2\over 2}
\int_a^b d z^\prime{\rho(z^\prime)\over z+z^\prime}\;=\;
{z\over 4\beta} - {d\over 2} - {d-2\over 2} f(-z)\;.
\label{s4e8}
\end{equation}
It follows that $R(z)$ is itself an analytic function of $z$,
with a cut on the negative real axis in the interval $[-b,-a]$.

Let us now notice that the normalization condition implies 
\begin{equation}
f(z) \goto_{|z|\rightarrow\infty}\; {1\over z}\;.
\label{s4e9}
\end{equation}
As a consequence in the same limit we obtain
\begin{equation}
R(z)\goto {z\over 4\beta} - {d\over 2} + {d-2\over 2z}
\label{s4e10}
\end{equation}
and in turn
\begin{equation}
\rho(z)\;=\; {1\over i\pi} \left[ R(z)-f(z)\right]\goto
{1\over i\pi}\left( {z\over 4\beta} - {d\over 2} + {d-4\over 2z}\right)
\,+\, O\left( {1\over z^2}\right)\;.
\label{s4e11}
\end{equation}
This equation can in principle be used in order to determine
relationships between the constants $a$ and $b$ in place of the
normalization condition.

We must now distinguish between weak and strong coupling regimes. In
both cases, by exploiting analyticity properties of the function
$f(z)$ and defining appropriate auxiliary functions, it is relatively
easy to reduce Eq.~(\ref{s4e3}) to the following forms
\begin{equation}
\rho(z) \;=\; {\sqrt{(b-z)(z-a)}\over \pi}\left[
{1\over 4\beta} - {d-2\over 2}\int_a^b {d y\over y+z}
{\rho(y)\over \sqrt{(b+y)(y+a)}}\right]
\;\;\;\;\;\;\;\;\;\;{\rm for}\;\;\;\; \beta > \beta_c\;,
\label{s4e12a}
\end{equation}
\begin{equation}
\rho(z) \;=\; {z\over \pi}\sqrt{b-z\over z-a}\left[
{1\over 4\beta} - {d-2\over 2}\int_a^b{d y\over y+z}
\sqrt{y+a\over y+b}{\rho(y)\over y}\right]
\;\;\;\;\;\;\;\;\;\;{\rm for}\;\;\;\; \beta < \beta_c\;.
\label{s4e12b}
\end{equation}
The values of $a$ and $b$ as functions of $\beta$ are determined by
enforcing the asymptotic condition (\ref{s4e11}).
Eqs.~(\ref{s4e12a}-\ref{s4e12b}) are inhomogeneous Fredholm equations
of the second kind.  It is therefore in principle possible to apply
standard methods of (approximate) resolution by expressing the kernels
in terms of appropriate orthonormal sets of eigenfunctions.

The weak and strong coupling constraints, Eq.~(\ref{wcond}) and
Eq.~(\ref{cond}), can be expressed in terms of the analytic function
$f(z)$ as $f(0)\geq -1$ and $f(0)=-1$ respectively.  Alternatively,
writing $f(z)=R(z) -i\pi \rho(z)$ and analytically continue this
expression outside of the interval $[a,b]$, Eq.~(\ref{wcond}) and
Eq.~(\ref{cond}), can also be expressed as
\begin{equation}
-i\rho(0)\;\geq \;0\;,
\label{s4e16}
\end{equation}
and 
\begin{equation}
-i\rho(0)\;= \;0\;,
\label{s4e16strong}
\end{equation}
respectively. Note that Eq.~(\ref{s4e12b}) is parameterized so that
the strong coupling constraint (\ref{s4e16strong}) is automatically
satisfied.  The transition point $\beta_c$ can be determined
approaching from the weak coupling regime by enforcing the equality
$\rho(0)=0$. We shall return to a general discussion of this
criticality in Section IV.

 A final comment concerns the explicit evaluation of
$\widetilde{Z}_d$. Instead of directly substituting $\rho(z)$ in the
expression of the partition function, it is convenient to apply
Eq.~(\ref{enem}) in the form
\begin{equation}
d(d-1)\,U\;=\;{1\over 4\beta^2}\int_a^b dz^\prime \, 
\rho(z^\prime)(z^{\prime\,2}-r)-d-{1\over\beta}\;,
\label{s4e14}
\end{equation}
and perform an integration with respect to $\beta$ to recover the free
energy.

\section{Exact Large N solutions}
%Exactly solvable examples.}
\label{sec5}

Here we present the solutions at $N = \infty$ as a function of
$d$. They can be broken into three classes for $\lowercase{d} < 4$,
$\lowercase{d} = 4$ and $\lowercase{d} > 4$ respectively. For
$\lowercase{d} < 4$, they are equivalent to chiral chain models with
$L<4$ studied earlier\cite{BRT} all of which exhibit the third order
Gross-Witten transition at $\beta_c$. For $\lowercase{d} >4$ there is
a first order transition, which ends exactly at $\lowercase{d} = 4$,
consequently the end point at $\lowercase{d} = 4$ is of special
interest.

\subsection{Solutions for $\lowercase{d} < 4$}

As we mentioned in the introduction, when $d$ is integer and less than
4 simplicial chiral models are only reformulation of trivial or
already solved models.  It is however quite instructive to consider
even these examples in our new language.  Let us begin with the only
apparently trivial case $d=0$. Obviously $Z_0=1$, however
$\widetilde{Z}_0$ is nontrivial and we need to know its value in order
to compute $Z_d$.  As a matter of fact Eq.~(\ref{tzm2}) already
implies that, up to a constant
\begin{equation}
\widetilde{Z}_0\;\propto\;\exp N^2\ln\beta\;.
\label{s5e1}
\end{equation}
We would like, as a consistency check, to derive this result
from the saddle point equation.
A straightforward manipulation of Eq.~(\ref{s4e3}) leads to
\begin{equation}
{1\over 8\beta}\;=\;\princint_a^b dz^\prime
{\rho(z^\prime)\over z^2-z^{\prime\,2}}\;.
\label{s5e2}
\end{equation}
This equation is solved by
\begin{equation}
\rho(z)\;=\; {z\over 4\pi\beta} {\sqrt{b^2-z^2}\over \sqrt{z^2-a^2}}\; ,
\label{s5e3}
\end{equation}
with the only constraint $b^2-a^2=16\beta$.  However by keeping in
mind that only the combination $x=z^2-a^2$ is physically meaningful
because of Eq.~(\ref{s4e1}), we recognize that
\begin{equation}
\rho(z)dz\;=\; {dz^2\over 8\pi\beta}{\sqrt{16\beta -(z^2-a^2)}\over
\sqrt{z^2-a^2}}\;=\; {1\over 8\pi\beta} {d x\over \sqrt{x}} \sqrt{16\beta-x}\; ,
\label{s5e4}
\end{equation}
and the physical solution is unique and leads by a trivial
integration, to Eq.~(\ref{s5e1}).

Because of our definitions $Z_1=1$, $\widetilde{Z}_1=\widetilde{Z}_0$.
The eigenvalue distribution $\rho(z)$ however is the generating
function for the moments of the linear combination of a complex and a
unitary matrix, and these moments can be highly nontrivial, even if
the complex matrix itself has a Gaussian probability distribution, as
a consequence of the averaging over unitary matrices.  As a matter of
fact we were not able to solve explicitly the saddle-point equation
associated to the $d=1$ models, even though the solution
probably has reasonably simple mathematical properties.

Let us now turn to the $d=2$ case.
As a straightforward application of Eqs.~(\ref{s4e12a}-\ref{s4e12b}),
we immediately find both the weak and strong coupling solutions,
\begin{equation}
\rho_{\rm w}(z)\;=\; {1\over 4\pi\beta}\sqrt{8\beta-(z-4\beta)^2}
\;\;\;\;\;\;\;\;\;{\rm for}\;\;\;\;\beta\geq {1\over 2}\;,
\label{s5e5a}
\end{equation}
\begin{equation}
\rho_{\rm s}(z)\;=\; {z\over 4\pi\beta}\sqrt{ 
{1+6\beta -z\over z-1+2\beta}}
\;\;\;\;\;\;\;\;\;{\rm for}\;\;\;\;\beta\leq {1\over 2}\;.
\label{s5e5b}
\end{equation}
For the strong coupling region, $r=(1-2\beta)^2$.

It is easy to recognize that the $d=2$ model corresponds to the 
Gross-Witten single-link problem, which in turn is equivalent to
large-$N$ QCD$_2$ with Wilson action on the lattice~\cite{Gross}.  The
properties of this model are well known, and in particular it is known
that $\beta_c=1/2$, consistent with
Eqs.~(\ref{s5e5a}-\ref{s5e5b}). Another consistency check is easily made by
applying Eq.~(\ref{s4e14}) and verifying that the known expressions
for $U_2$ are reproduced.

Finally let us comment about the $d=3$ case. This model in its
original formulation is completely equivalent to the three-link chiral
chain studied in Refs.~\cite{BRT,Friedan}.  We therefore know that it
must possess a third-order phase transition at  the
critical value $\beta_c=1/3$. However, as we already observed, our
reformulation leads to exploring quite different classes of
correlation functions and there is no obvious relationship between old
and new results apart from bulk thermodynamical properties.  Again we
have no analytical solution for the $d=3$ model equation, whose known
properties stand as a benchmark for future attempts.

\subsection{Solution at  $\lowercase{d}=4$}
\label{sec6}

Turning to $d=4$ leads us to a new situation, where we are no longer
guided by known results, since the 3-dimensional simplex (tetrahedron)
is distinct from the solved four-link chain. Actually it would be
instructive and convenient to embed both models in a more general case
interpolating between them and including many more interesting
situations. We are studying  the most general four-site system
with bilinear interactions of four unitary matrices, which turns
out to be reducible to an interacting two-complex matrix system.  A
separate paper will be devoted to a discussion of this system.  Here
we only discuss the solutions of the saddle point equation obtained
from Eq.~(\ref{s4e3}) in the $d=4$ case,
\begin{equation}
{z\over 8\beta}-1\;=\;\princint d z^\prime{z'\rho(z^\prime)\over
z^2-z^{\prime\,2}}\;.
\label{s6e7}
\end{equation}

In order to solve this equation, let us separately consider the weak and the
strong coupling regimes, while changing variables for convenience to $x=z^2$ 
and defining the distribution $\tilde{\rho}(x)$ by 
$\rho(z)dz=\tilde{\rho}(x)dx$.
The special structure of Eq.~(\ref{s6e7}) makes it convenient to follow a special
procedure not directly related to Eqs.~(\ref{s4e12a}-\ref{s4e12b})
derived for the general case.

In the weak coupling phase, we define the functions 
\begin{equation}
f_{\rm w} (x) \;=\; \int_{a^2}^{b^2} dx^\prime{\tilde{\rho}_{\rm w}
(x^\prime) \sqrt{x'}\over x-x^\prime}\; ,
\label{s6e8}
\end{equation}
and
\begin{equation}
g_{\rm w}(x)\;=\; {f_{\rm w}(x)\over \sqrt{(x-b^2)(x-a^2)}}\;,
\label{s6e9}
\end{equation}
subject to the normalization constraint
\begin{equation}
\int_{a^2}^{b^2}dx^\prime\,\tilde{\rho}_{\rm w}(x^\prime)\;=\;1\;.
\label{s6e10}
\end{equation}
The functions $f(x)$ and $g(x)$ are real analytic, with a cut along
the interval $[a^2,b^2]$ on the real axis. On this interval the relationship,
\begin{equation}
{\rm Im} \,g_{\rm w}(x\pm i\epsilon)\;=\;\mp {{\rm Re}\,f_{\rm w}(x)\over
\sqrt{(x-a^2)(b^2-x)}}\;,
\label{s6e11}
\end{equation}
holds, while analyticity and Eq.~(\ref{s6e7}) imply
\begin{equation}
g_{\rm w}(x)\;=\;
{ {\sqrt{x}\over 8\beta}-1\over \sqrt{(x-b^2)(x-a^2)}}
-{1\over 8\pi\beta}\int_0^\infty dy\,
{\sqrt{y}\over x+y}{1\over \sqrt{ (y+b^2)(y+a^2)}}
\;.
\label{s6e12}
\end{equation}
However Eqs.~(\ref{s6e8}-\ref{s6e9}) imply that
\begin{equation}
\tilde{\rho}_{\rm w}(x)\;=\; -{{\rm Im}\,f_{\rm w}(x+i\epsilon)\over \pi\sqrt{x}}
\;=\; {1\over 8\pi^2\beta}\int_0^\infty {d y\over x+y}
\sqrt{{y\over x}} \sqrt{ {(b^2-x)(x-a^2)\over (b^2+y)(y+a^2)}}\;.
\label{s6e13}
\end{equation}
In order to determine $a^2$ and $b^2$ we may use Eq.~(\ref{s6e10})
and the observation that in the complex $x$ plane when $|x|\rightarrow \infty$
\begin{equation}
g_{\rm w}(x)\goto O\left({1\over x^2}\right)\;.
\label{s6e14}
\end{equation}
As a consequence one obtains that
\begin{equation}
\oint dx^\prime\,g_{\rm w}(x^\prime)\;=\; 2\int_{a^2}^{b^2}
dx^\prime\, {\rm Im}\,g_{\rm w}(x^\prime)\;=\;0
\label{s6e15}
\end{equation}
around $[a^2,b^2]$, that is
\begin{equation}
\int_{a^2}^{b^2} dx^\prime 
{\sqrt{x^\prime}\over \sqrt{(b^2-x^\prime)(x^\prime-a^2)}}\;=\; 8\pi\beta\;.
\label{s6e16}
\end{equation}
In strong coupling we adopt a similar strategy by defining 
\begin{equation}
f_{\rm s} (x)\;=\; \int_{a^2}^{b^2} {dx'\over \sqrt{x'}}
{\tilde{\rho}_{\rm s}(x')\over x-x'}\;,
\label{s6e17}
\end{equation}
and
\begin{equation}
g_{\rm s}(x)\;=\; f_{\rm s}(x)\sqrt{ {x-a^2\over x-b^2}}\;,
\label{s6e18}
\end{equation}
with the constraint
\begin{equation}
\int_{a^2}^{b^2}\tilde{\rho}_{\rm s}(x') d x'\;=\;1\;,
\label{s6e19}
\end{equation}
and the boundary condition
\begin{equation}
g_{\rm s}(x)\goto_{|x|\rightarrow \infty} {1\over x}\;.
\label{s6e20}
\end{equation}
We then find
\begin{equation}
g_{\rm s}(x)\;=\; {1\over 8\beta\sqrt{x}}\sqrt{ {x-a^2\over x-b^2}}
-{1\over 8\pi\beta}\int {dy \over x+y} {1\over \sqrt{y}}
\sqrt{ {y+a^2\over y+b^2}}\;,
\label{s6e21}
\end{equation}
and
\begin{equation}
\tilde{\rho}_{\rm s}(x)\;=\;{1\over 8\pi^2\beta}
\int_{a^2}^{b^2} { dy\over x+y} 
\sqrt{ {x(b^2-x)(y+a^2)\over y(b^2+y)(x-a^2)}}\;.
\label{s6e22}
\end{equation}
The boundary condition (\ref{s6e20}) leads to the relationship
\begin{equation}
-\int_{a^2}^{b^2}dx' \,{\rm Im}\,g_{\rm s}(x'+i\epsilon)\;=\;\pi\;,
\label{s6e23}
\end{equation}
where
\begin{equation}
{\rm Im}\,g_{\rm s}(x+i\epsilon)\;=\;- {1\over 8\beta\sqrt{x}}
\sqrt{ {x-a^2\over b^2-x}}\;.
\label{s6e23b}
\end{equation}

All the integrals appearing in Eqs.~(\ref{s6e13}), (\ref{s6e16}),
(\ref{s6e22}) and (\ref{s6e23}) are elliptic integrals. It is
therefore possible to re-express both the weak and the strong coupling
results in terms of known functions. In particular it is convenient to
re-express everything in terms of ``natural'' rescaled variables, by
defining
\begin{equation}
k\;=\;\sqrt{ 1 - {a^2\over b^2}}\;,
\label{s6e24}
\end{equation}
\begin{equation}
\zeta\;=\;\sqrt{ 1-{z^2\over b^2}}\;,
\label{s6e25}
\end{equation}
and setting $\bar{\rho}(\zeta)d\zeta=\rho(z)dz$.
It is not too difficult to eliminate completely the parameters $a$ and $b$
in favor of $k$ by making use of Eqs.~(\ref{s6e16}) and (\ref{s6e23})
respectively.
As a consequence we obtain the weak coupling expression
\begin{equation}
\bar{\rho}_{\rm w}(\zeta)\;=\;
{8\beta\over E(k)^2}\left[ {\sqrt{k^2-\zeta^2}\over \sqrt{1-\zeta^2}}
K(k) - \sqrt{k^2-\zeta^2}\sqrt{1-\zeta^2}\Pi(\zeta^2,k)\right]\;,
\label{s6e26}
\end{equation}
and the strong coupling counterpart
\begin{equation}
\bar{\rho}_{\rm s}(\zeta)\;=\;
{8\beta\over \left[ E(k)-(1-k^2)K(k)\right]^2}
\left[ k^2{\sqrt{1-\zeta^2}\over \sqrt{k^2-\zeta^2}}K(k)
-\sqrt{k^2-\zeta^2}\sqrt{1-\zeta^2} \Pi(\zeta^2,k)\right]\;,
\label{s6e27}
\end{equation}
where $K,E,\Pi$ are the elliptic integrals of the first, second and third kind 
respectively, and the domain of $\zeta$ is the interval $[0,k]$, $0\leq k\leq 1$.

Obviously, in order for the problem to be completely solved, one must
try expressing $k$ as a function of $\beta$. This is achieved in principle by 
enforcing the normalization condition, which takes the form
\begin{equation}
\int_0^k \bar{\rho}(\zeta)d\zeta\;=\;1\;.
\label{s6e28}
\end{equation}
By symbolically writing
\begin{equation}
\bar{\rho}(\zeta)\;=\;\beta D(\zeta,k)\;,
\label{s6e29a}
\end{equation}
in agreement with Eqs.~(\ref{s6e26}) and (\ref{s6e27}), it is actually possible
to express all results as functions of $k$ by the relationship
\begin{equation}
\beta\;=\; {1\over \int_0^k d\zeta\,D(\zeta,k)}\;.
\label{s6e29}
\end{equation}

In practice this form of our results is sufficient for both numerical
evaluation and asymptotic expansions, not to mention the possibility
of exploring the region around the criticality.  Criticality is
characterized by the limit $k\rightarrow 1$, where simple mathematical
properties of elliptic integrals allow us to show that both weak and
strong coupling results lead to $\beta_c=1/4$ and
\begin{equation}
\bar{\rho}_c(\zeta)\;=\;\zeta\ln {1+\zeta\over 1-\zeta}\;.
\label{s6e30}
\end{equation}

In order to obtain the usual weak and strong coupling expansion
of physical quantities, like the internal energy, as power series in $1/\beta$
and $\beta$ respectively, one must consider in turn the $k\rightarrow 0$
limit and the expansion in powers of $k$. Obviously
the different structure of $\bar{\rho}(\zeta)$ 
in the two phases will lead to different expressions.
In particular we have the asymptotic behaviors
\begin{equation}
\beta_{\rm w}\goto_{k\rightarrow 0}
{2\over k^4} - {2\over k^2} + O(1)\;,
\label{s6e31a}
\end{equation}
\begin{equation}
\beta_{\rm s}\goto_{k\rightarrow 0} {k^2\over 16} + {k^4\over 32}+O(k^6)\;,
\label{s6e31b}
\end{equation}
and it is conceptually straightforward to obtain power series
expansions in the powers of $k$ for such quantities as the internal
energy and to convert them into standard weak and strong coupling
series. A few details will be discussed in Appendix~\ref{appc}.

The expansion around the critical point $\beta_c=1/4$, $k_c=1$, is
slightly subtler because the expansion of elliptic integrals around
$k=1$ is asymptotic.  However by exploiting a few known or previously
derived results, we have managed to obtain the following
relationships, holding in weak coupling near the criticality:
\begin{equation}
\bar{\rho}_{\rm w}(\zeta)\;\approx\;
4\beta\left[  1-k'^2\left(\ln{4\over k'}-{1\over 2}\right)+O(k'^2)\right]
\left[ \zeta \ln {1+\zeta\over 1-\zeta}-
k'^2\left( \ln {4\over k'} -{1\over 2}\right){\zeta^2\over 1-\zeta^2}
+O(k'^2)\right]\;,
\label{s6e32}
\end{equation}
where $k'^2=1-k^2\rightarrow 0$.
From Eq.~(\ref{s6e29}) we then obtain
\begin{equation}
4\beta\;\approx \; 1+k'^2
\left[
\ln{4\over k'}\ln{2\over k'}+{1\over 2}\ln{2\over k'}+{1\over 2}\right]\;,
\label{s6e33}
\end{equation}
and as a consequence
\begin{equation}
\beta-\beta_c\;\approx\;{k'^2\over 4}\left[
\ln{4\over k'}\ln{2\over k'}+{1\over 2}\ln{2\over k'}+{1\over 2}\right]\;,
\label{s6e34}
\end{equation}
therefore $k'^2\sim \beta-\beta_c$ apart from logarithms.
By properly applying Eq.~(\ref{s4e14}) we may also extract the result
\begin{equation}
U_{\rm w}\goto_{k'\rightarrow 0}
{\pi^2-6\over 9}+O(k'^2\ln^2 k')\;,
\label{s6e35}
\end{equation}
and by simple manipulations, from the specific heat relationship
\begin{equation}
C\;=\;\beta^2{\partial U\over\partial\beta}\;,
\label{s6e36}
\end{equation}
we may obtain near the criticality
\begin{equation}
C_{\rm w}\goto_{k'\rightarrow 0}
{\pi^2+3\over 36}-{\pi^2\over 12 \ln(4/k')} 
+O\left( {1\over \ln^2 k'}\right)\;.
\label{s6e37}
\end{equation}
A similar analysis can be performed in strong coupling near criticality,
\begin{equation}
\bar{\rho_{\rm s}}(\zeta)\;\approx\;
4\beta\left[  1+k'^2\left(\ln{4\over k'}+{1\over 2}\right)+O(k'^2)\right]
\left[ \zeta \ln {1+\zeta\over 1-\zeta}+
k'^2\left( \ln {4\over k'} +{1\over 2}\right){\zeta^2\over 1-\zeta^2}
+O(k'^2)\right]\;,
\label{s6e38}
\end{equation}
where again $k'^2=1-k^2\rightarrow 0$, and
\begin{equation}
\beta_c-\beta\;\approx\;{k'^2\over 4}\left[
\ln{4\over k'}\ln{2\over k'}-{1\over 2}\ln{2\over k'}-{1\over 2}\right]\;.
\label{s6e39}
\end{equation}
We then find
\begin{equation}
C_{\rm s}\goto_{k'\rightarrow 0}
{\pi^2+3\over 36}-{\pi^2\over 12 \ln(4/k')} 
+O\left( {1\over \ln^2 k'}\right)\;.
\label{s6e40}
\end{equation}
The strong and weak coupling expressions of
$C$ near the criticality  show that the critical behavior around
$\beta_c={1\over 4}$ corresponds to a limiting case of a third order
phase transition with critical exponent of the specific heat
\begin{equation}
\alpha\;=\;0^-\;,
\label{s6e41}
\end{equation}
near the boundary with weak second order critical behavior.
Notice that in terms of double scaling limit $\alpha=0^-$ would correspond to
a central charge $c=1$.

For the interested readers we mention that in the derivation of Eqs.~(\ref{s6e32})
and (\ref{s6e38}) we made use of the following formula (which 
appeared with some misprints in Ref.~\cite{pi})
\begin{equation}
\Pi(\zeta^2,k)\;\approx\; 
{1\over 1-\zeta^2}\left( \ln{4\over k'}+{\zeta\over 2}
\ln{1-\zeta\over 1+\zeta}\right) + 
{k'^2\over 4(1-\zeta^2)^2}
\left(-1+(1+\zeta^2)\ln{4\over k'} +
\zeta\ln{1-\zeta\over 1+\zeta}\right)\;,
\label{s6e42}
\end{equation}
for the
asymptotic expansion of the elliptic integral of the third kind $\Pi(\zeta^2,k)$
in the region $k'=\sqrt{1-k^2}\rightarrow 0$.

\subsection{The  $\lowercase{d}=\infty$ solution}
\label{sec8}

While at present we are not aware of any general method to get an
analytic solution of the saddle-point equation (\ref{s4e3}) for
arbitrary $d$, the $d\rightarrow\infty$ limit provides another
interesting instance in which the equation is solvable.

It is easy to show that for larger and larger values of $d$ the
distribution $\rho(z)$ becomes narrower and narrower, with a width
decreasing like $d^{-1/2}$ and a peak value $\bar{z}$ which can easily
be determined by replacing in Eq.~(\ref{s4e3})
\begin{equation}
\rho(z)\goto \delta(z-\bar{z})\;,
\label{s8e1}
\end{equation}
and obtaining the large-$N$, large-$d$ equation
\begin{equation}
{\bar{z}\over 2\beta}-d\;=\;-{d-2\over 2\bar{z}}\;.
\label{s8e2}
\end{equation}
A consistent solution is obtained by assuming the limit to be taken
at a fixed value of $\beta d$, in which case 
\begin{equation}
\bar{z}\goto \beta d\left( 1 + \sqrt{1-{1\over \beta d}}\right)  k
\label{s8e3}
\end{equation}
with the obvious restriction $\beta d \geq 1$ (weak coupling phase).
When $\beta d \leq 1$ one must recognize that the saddle-point condition,
when correctly applied to the original expression for the effective action 
Eq.~(\ref{seff})  in the large-$d$ limit, unambiguously leads to the prediction 
$\bar{z}\rightarrow 1$, $r\rightarrow 1$ (strong coupling phase).

The most interesting features of this result are:
\begin{itemize}
\item The large-$d$ prediction for the location of the critical point, 
$\beta_c\goto 1/d$, amazingly enough, seems to be
satisfied for all values of $d$.

\item The complete equivalence with the mean field solution of infinite volume
principal chiral models on a $D$-dimensional hypercubic lattice such
that $D=d/2$~\cite{Kogut}, where we may observe that this last
relationship enforces the constraint that corresponding models have
the same coordination number.
\end{itemize}

It is easy to compute the large-$d$ expression for the internal energy
in the weak coupling,
\begin{equation}
U\goto {1\over 4(\beta d)^2}\bar{z}^2\;=\;
{1\over 2} + {1\over 2}\sqrt{1-{1\over \beta d}}-{1\over 4\beta d}\;.
\label{s8e5}
\end{equation}
  At the criticality. the weak coupling value of $U$
is ${1\over 4}$, while the strong coupling value is $U=0$.  Therefore
the large-$N$, large-$d$ prediction for the nature of the criticality
is that of a first-order phase transition.  It is however important to
notice that the large-$d$ prediction for the specific heat in the
weak-coupling phase,
\begin{equation}
dC\;=\; {1\over 4} \left[ {1\over \sqrt{ 1 - {1\over \beta d}}}+1\right]\;,
\label{s8e5b}
\end{equation}
shows a divergence at the phase
transition, with no indication for the existence of a metastable
phase. 

It is interesting to compare 
the specific heat behavior for $d=2,3,4,\infty$.
In Fig.~\ref{dC} we plotted $dC$ versus $\bar{\beta}\equiv d\beta$.

\subsection{The  $1/\lowercase{d}$ expansion}
\label{sec8a}

The large-$d$ result may also be the starting point for a systematic
$1/d$ expansion of Eq.~(\ref{s4e3}), and for a numerical approximation
scheme which turns out to be quite efficient at least in the weak coupling domain
away from criticality. The essential ingredient for both these developments is 
the observation that, substituting the definition of $f(z)$, Eq.~(\ref{s4e6}),
into Eq.~(\ref{s4e3}) we obtain the functional equation,
\begin{equation}
{z\over 2\beta}-d\;=\; 2{\rm Re} \,f(z)+(d-2)f(-z)\; ,
\label{s8e6}
\end{equation}
subject to the following constraints: (a) Eq.~(\ref{s8e6}) is
satisfied in the interval $[a,b]$, with $a$ and $b$ dynamically
determined; (b) $f(z)$ is real analytic outside the interval $[a,b]$;
(c) the asymptotic behavior of $f(z)$ when $|z|\rightarrow \infty$ is
$f(z)\goto {1/ z}$.

Let us now define
\begin{equation}
\zeta \equiv  z-\tilde{z}  \;  , \; \; \; \; \phi(\zeta) \equiv f(\tilde{z}+\zeta)\;,
\label{s8e8}
\end{equation}
where $\tilde{z}$ is a constant whose value lies in the interval
$[a,b]$ and will be dynamically generated. Substituting the definition
(\ref{s8e8}) into Eq.~(\ref{s8e6}), we obtain
\begin{equation}
{\tilde{z}\over 2\beta}-d+{\zeta\over 2\beta}\;=\;
2{\rm Re}\,\phi(\zeta)+(d-2)\phi(-2\tilde{z}-\zeta)\;.
\label{s8e9}
\end{equation}
Postponing the discussion of the numerical approximation scheme, let
us illustrate here the procedure for a systematic $1/d$ expansion of
Eq.~(\ref{s8e9}).  We introduce the following Ansatz for the function
$\phi(\zeta)$,
\begin{equation}
\phi(\zeta)\;=\; d Q(\zeta)\left( 1-\sqrt{1-{c_1\over d\zeta}-{c_2\over d\zeta^2}}\right)
+R(\zeta)\;,
\label{s8e10}
\end{equation}
and assume the functions $Q(\zeta)$, $R(\zeta)$ to be real analytic in the interval
between the roots of the polynomial
\begin{equation}
d\zeta^2-c_1\zeta-c_2\;,
\end{equation}
including the point $\zeta=0$. Moreover we require $Q(0)=R(0)=0$ and the boundary
condition 
\begin{equation}
\phi(\zeta)\goto_{|\zeta|\rightarrow \infty}{1\over \zeta}\;.
\end{equation}
These requirements fix the constants $\tilde{z}$, $c_1$ and $c_2$ dynamically,
which in turn determine $a$ and $b$.
Finally we assume all functions and constants to be expandable in $1/d$, 
with non vanishing leading order.

As an illustration let us find the solution to first nontrivial
order. All leading order quantities will be independent of $d$ and
labeled by a subscript 0.  After an expansion of Eq.~(\ref{s8e10}) in
powers of $1/d$ we obtain
\begin{equation}
\phi(\zeta)\;\approx\; 
{1\over 2}Q_0(\zeta)\left(
{c_1\over \zeta}+{c_2\over \zeta^2}\right) + R_0(\zeta)+
O\left( {1\over d}\right)\;.
\label{s8e11}
\end{equation}
By imposing the asymptotic boundary conditions we have
the stricter condition (forced by analyticity of $Q_0$, $R_0$),
\begin{equation}
{1\over 2}Q_0(\zeta)\left(
{c_1\over \zeta}+{c_2\over \zeta^2}\right) + R_0(\zeta)\;=\;{1\over \zeta}\;.
\label{s8e12}
\end{equation}
As a consequence we may also predict the $O\left({1\over d}\right)$
asymptotic behavior,
\begin{equation}
\phi(\zeta)\goto {1\over \zeta} + {1\over d}{1\over 4\zeta^2}
\left( c_1+{c_2\over \zeta}\right) + O\left({1\over d^2}\right)\;.
\label{s8e13}
\end{equation}
Now by substituting the above results into Eq.~(\ref{s8e9})
we obtain, after expansion in powers of $1/d$,
\begin{equation}
{\tilde{z}_0+\zeta\over 2\beta d}-1\;=\;
2Q_0(\zeta)-{1\over 2\tilde{z}_0+\zeta}\;,
\label{s8e14}
\end{equation}
where we always assume the large-$d$ limit to be taken while keeping $\beta d$ finite.
Substituting the condition $Q_0(0)=0$ into Eq.~(\ref{s8e14})
we may solve it in the form,
\begin{equation}
{1\over 2\beta d}\;=\; {1\over \tilde{z}_0}-{1\over 2\tilde{z}_0^2}\;,
\label{s8e15a}
\end{equation}
\begin{equation}
Q_0(\zeta)\;=\; {\zeta\over 2\tilde{z}_0}\left[ \left(1-{3\over 4\tilde{z}_0}\right)
+{1\over 4\tilde{z}_0}{\zeta\over 2\tilde{z}_0+\zeta}\right]\;,
\label{s8e15b}
\end{equation}
while the implementation of Eq.~(\ref{s8e12}) fixes
\begin{eqnarray}
c_1&=&-{1\over 2\tilde{z}_0} {1\over \left( 1-{3\over 4\tilde{z}_0}\right)^2}\;,
\nonumber \\
c_2&=&{4\tilde{z}_0\over \left( 1-{3\over 4\tilde{z}_0}\right)}\;,
\label{s8e16}
\end{eqnarray}
and in conclusion we may also write
\begin{equation}
R_0(\zeta)\;=\;
{\left( 1-{1\over 2\tilde{z}_0}\right)\over 
8\tilde{z}_0^2\left( 1 - {3\over 4\tilde{z}_0}\right)^2}
{\zeta\over 2\tilde{z}_0+\zeta}\;.
\label{s8e17}
\end{equation}
Let us notice that the leading order is completely determined in terms
of the parameter $\tilde{z}_0$, which in turn is fixed through
Eq.~(\ref{s8e15a}) to take the value
\begin{equation}
\tilde{z}_0\;=\;\bar{z}\;=\;\beta d\left( 1 +\sqrt{1-{1\over \beta d}}\right)\;.
\label{s8e18}
\end{equation}
Hence we recognize that $\tilde{z}$ is nothing but a generalization of
the mean field parameter which (roughly speaking) describes the center
of the eigenvalue distribution, while the width of the distribution
itself is $O\left(1/\sqrt{d}\right)$ as one may easily see by studying
the roots of the polynomial under the square root sign.

The eigenvalue distribution itself may be recovered (order by order in $1/d$)
by the relationship
\begin{equation}
\rho(z)\;=\; 
{d Q(z-\tilde{z})\over \pi (z-\tilde{z})}
\sqrt{ {c_2\over d} + {c_1\over d}(z-\tilde{z}) - (z-\tilde{z})^2}
\;,
\label{s8e19}
\end{equation}
and it is not too difficult to check that the large-$d$ limit of Eq.~(\ref{s8e19})
may be taken and the result is
\begin{equation}
\rho(z)\goto \lim_{d\rightarrow\infty}
{d\over \pi} {2\over c_2}\sqrt{{c_2\over d}-(z-\tilde{z}_0)^2}\;=\;\delta(z-\tilde{z}_0)\;,
\label{s8e20}
\end{equation}
as expected.
We have worked out higher orders of the $1/d$ expansion. 

The Ansatz (\ref{s8e10}) can also be used as a starting point for numerical
approximations, and this will be described in Appendix D. For instance, with the
precision of about $1\%$, one can quickly verify numerically  the conjecture that
$\beta_c=1/d$.

\section{Simplicial models at criticality}
\label{sec7}

We would like to be able to understand in more detail the behavior of
the critical properties as a function of d. Eq.~(\ref{s4e3}), to the
best of our knowledge, does not lend itself to an exact treatment for
arbitrary values of $\beta$ and $d$. However, on the critical surface for
weak-strong transition, we find that it is possible to turn
Eqs.~(\ref{s4e12a})-(\ref{s4e12b}) into a homogeneous (eigenvalue)
equation. For $\lowercase{d} \leq 4$, this eigenvalue problem can be
solved analytically. For $\lowercase{d} > 4$, the problem can be
solved numerically with great precision.

It is also worth pointing out that at criticality for $\lowercase{d}
\leq 4$, Eq.~(\ref{s4e3}) can be solved by applying a method of Gaudin
and Kostov for the study of $O(n)$ spins on random surfaces. There is in
fact an exact mapping of our weak-coupling  critical saddle point equation to
that of Ref.~\cite{Gaudin}, with
$d=n+2$. However, for $\lowercase{d} > 4$, the Gaudin-Kostov's
solution become pathological. In contrast, for the simplicial models,
we find that a consistent solution exists for $\lowercase{d} >
4$.

The determination of $\beta_c$ can be achieved by considering
Eq.~(\ref{s4e12a}) in the limit when the equality in the weak coupling
constraint, Eq.~(\ref{s4e16}), is reached.  On the weak coupling side
of criticality the condition $\rho(0)=0$ implies
\begin{equation}
\sqrt{a_cb_c}\left[ {1\over 4\beta_c}-{d-2\over 2}
\int_{a_c}^{b_c} {dy\over y} {\rho_c(y)\over \sqrt{(b_c+y)(y+a_c)}}\right]
\;=\;0\;.
\label{s4e17}
\end{equation}
There are two possible solutions, (i) $a_c=0$, and (ii) $a_c\neq 0$, with 
\begin{equation}
{1\over \beta_c}\;=\; 2(d-2)\int_{a_c}^{b_c} {dy\over y}
{\rho_c(y)\over \sqrt{(b_c+y)(y+a_c)}}\;.
\label{s4e19}
\end{equation}
If Eq.~(\ref{s4e17}) is solved by $a_c=0$ at $\beta=\beta_c$,
Eqs.~(\ref{s4e12a}) and (\ref{s4e12b}) both reduce to
\begin{equation}
\rho_c(z) \;=\; {\sqrt{(b_c-z)z}\over \pi}\left[
{1\over 4\beta_c} - {d-2\over 2}\int_0^{b_c} {d y\over y+z}
{\rho_c(y)\over \sqrt{(b_c+y)y}}\right]\;.
\label{s4e18}
\end{equation}
This indeed applies for $0\leq d\leq 4$, and we shall find an explicit
solution of Eq.~(\ref{s4e18}), which agrees with a result previously
found by Gaudin and Kostov~\cite{Gaudin}. In particular, we find that
$\beta_c=1/d$ for $0\leq d \leq 4$.

When $d>4$, Eq.~(\ref{s4e17}) must be solved with
$a_c>0$. Substituting Eq.~(\ref{s4e19}) into
Eqs.~(\ref{s4e12a})-(\ref{s4e12b}), one arrives at a homogeneous
equation
\begin{equation}
\lambda \rho_c(z)\;=\; z\sqrt{(b_c-z)(z-a_c)}\int_{a_c}^{b_c}
{dy\over y( y+z)}{\rho_c(y)\over \sqrt{(b_c+y)(y+a_c)}}\;,
\label{s4e20}
\end{equation}
where $\lambda={2\pi\over d-2}$. Eq.~(\ref{s4e20}) can be solved
numerically with great accuracy. Surprisingly, the relationship
$\beta_c=1/d$ was found to be satisfied within machine precision.  The
essential feature of the solution for $d>4$ is the strong coupling
relationship $r_c={a_c}^2\neq 0$, while in weak coupling necessarily
$r=0$.  Since $\rho_c(z)$ is the same on both sides of the transition,
at criticality one finds from Eq.~(\ref{s4e14})
\begin{equation}
U^{(c)}_{\rm w}\,-\,U^{(c)}_{\rm s}\;=\;
{1\over 4\beta_c^2}\int_{a_c}^{b_c}dz^\prime {\rho_c(z^\prime)r_c\over d(d-1)}\;=\;
{d\over 4(d-1)}a_c^2\;,
\label{s4e21}
\end{equation}
and as a consequence a first order phase transition is observed.

\subsection{Critical Solution for $0\leq d\leq 4$}

When $0\leq d\leq 4$, Eq.~(\ref{s4e3}) at criticality can be solved by
 assuming that $a_c=0$, as suggested by our analytic results discussed
 in Section III. Let us therefore focus on Eq.~(\ref{s4e18}). This
 equation on the first sight suggests that $\rho(z)$ would vanish at
 $z=0$ as $\sqrt z$. However, it is easy to verify that, upon
 substituting this behavior into the right hand side of the equation,
 this square-root behavior is in fact inconsistent. Based on our
 earlier exact analytic solutions, we assume that $\rho_c(z)$ vanishes
 at $z=0$ faster the $\sqrt z$; it follows that the square-bracket in
 Eq.~(\ref{s4e18}) must also vanish at $z=0$. As a consequence, we
 have
\begin{equation}
{1\over \beta_c}\;=\; 2(d-2)\int_{a_c}^{b_c} {dy\over y^{3/2}}
{\rho_c(y)\over \sqrt{b_c+y}}\;,
\label{s4e22}
\end{equation}
and we  again arrive at a homogeneous equation
\begin{equation}
\lambda \rho_c(z)\;=\; z^{3/2}\sqrt{b_c-z}\int_{0}^{b_c}
{dy\over y+z}{\rho_c(y)\over y^{3/2}\sqrt{b_c+y}}\;.
\label{s4e23}
\end{equation}
Note that this homogeneous equation connects smoothly with that
appropriate for $d>4$, Eq.~(\ref{s4e20}), with $a_c=0$.

It is convenient to change variable from $z$ to $\omega$,
$\omega+\omega^{-1}=2b_c/z$. In solving for $\omega$ in terms of $z$, we
shall choose the branch  $1\leq \omega <
\infty$ so that Eq.~(\ref{s4e23}) becomes 
\begin{equation}
\lambda f(\omega)\;=\; (\omega^{1/2}-\omega^{-1/2})\int_{1}^{\infty}
{d\omega'(\omega'^{1/2}-\omega'^{-1/2})\over(\omega'+\omega)
(\omega'+\omega^{-1})}{f(\omega')}\;,
\label{s7e26}
\end{equation}
where  $f(\omega)=(1/2)(\omega +\omega^{-1})\rho_c(z)$.

Although $f(\omega)$ is originally defined only for the interval
$1\leq \omega<\infty$, the right-hand side of Eq.~(\ref{s7e26})
provides a natural extension to the region $0\leq
\omega\leq 1$. With this extension, one finds that, over the positive axis, $0\leq
\omega<\infty$,  
\begin{equation}
f(\omega)\;=\; - f(\omega^{-1})\;.
\label{s7e27}
\end{equation}
It is then straightforward to verify that this extension can also be
made for the right-hand side of Eq.~(\ref{s7e26}) so that it becomes
\begin{equation}
\lambda f(\omega)\;=\; {(\omega^{1/2}-\omega^{-1/2})\over 2}\int_{0}^{\infty}
{d\omega'(\omega'^{1/2}-\omega'^{-1/2})\over(\omega'+\omega)
(\omega'+\omega^{-1})}{f(\omega')}\;.
\label{s7e28}
\end{equation}

Let us next treat Eq.~(\ref{s7e28}) as an eigenvalue problem, and
consider the Ansatz where
$f(\omega)=c[\omega^{\theta}-\omega^{-\theta}]$. Using the technique
of contour-integration, it is easy to verify that this indeed is an
eigenvector with eigenvalue $\lambda=\pi/\cos{\pi\theta}.$ Since
$\lambda = {2\pi\over {d-2}}$, it follows that $d\; =\; 4 {\rm
cos}^2{\pi\theta\over 2}$. With $0\leq d\leq 4$, one has $\theta$ real
and $0\leq \theta\leq 1$, which allows a solution where $\rho_c(z)$ is
positive definite!

Using the normalization condition for $\rho_c$ together with the criticality
condition $\int_0^{b_c}{dz'/ z'}\rho_c(z')=1$, we can fix the normalization
constant $c= {1\over
\pi \theta}{\rm cos} {\pi\theta\over 2}\;$ and the end point   
$b_c= {2\over
\theta}{\rm tg} {\pi\theta\over 2}\;.$ One then obtains
\begin{equation}
\rho_c(z)\;=\;{{\rm cos}{\pi\theta\over 2}\over {\pi\theta\over 2}}
{{\rm sh}\theta u\over {\rm ch} u}\;,
\label{s7e6}
\end{equation}
where $e^u=\omega$ and $0\leq u< \infty$. One can show that
Eq.~(\ref{s7e6}) reproduces the known critical solution when $d=2$ and
4. When substituted into the critical equation at $d=3$
Eq.~(\ref{s7e6}) is numerically found to be a satisfactory solution.

To determine the critical value $\beta_c$, we can  re-express
Eq.~(\ref{s4e22}) in terms of $f(\omega)$ as 
\begin{equation}
{1\over \beta_c}\;=\;{ (d-2)\over {\sqrt 2}b_c} \int_{0}^{\infty} d\omega
{{(\omega^{1/2}-\omega^{-1/2})}\over{1+\omega^2} } f(\omega)\;.
\label{s7e29}
\end{equation}
Again, by an contour integration, one arrives at the remarkable result
\begin{eqnarray}
\beta_c={1\over 4 {\rm cos}^2{\pi\theta\over 2}}\;=\;{1\over d} \; .
\label{s7e30}
\end{eqnarray}

\subsection{Criticality for $d > 4$}
\label{sec9a}

The solution discussed above does not apply to the case $d>4$ because
the analytic continuation of Eq.~(\ref{s7e6}) for the critical density
would no longer be positive-definite in the interval $[0,b_c]$; we
must choose the alternative, $a_c\neq 0$.  We have previously
seen, with $a_c\neq 0$, how the criticality condition for $\beta_c$,
Eq~(\ref{s4e19}), and the homogeneous integral equation for $\rho_c$,
Eq.~(\ref{s4e20}), can be obtained, approaching from the weak coupling
regime, by enforcing the condition $\rho(0)=0$ with $r=0$. It is
instructive to see how these equations can be similarly derived
from the strong coupling regime.

Starting with Eq.~(\ref{s4e12b}), one is working within the strong
coupling regime where the constraint, Eq.~(\ref{cond}), is
automatically satisfied with $a={\sqrt r}\neq 0$. It can be shown that
the criticality condition, Eq.~(\ref{s4e19}), corresponds to a
situation where a zero of $\rho(z)$ enters at $z=a$. That is, as one
increases $\beta$ beyond $\beta_c$, the positivity of $\rho(z)$ would
be violated, thus terminating the validity of the strong coupling
solution. As pointed earlier, with Eq.~(\ref{s4e19}), the strong
coupling equation, Eq.~(\ref{s4e12b}), again leads to
Eq.~(\ref{s4e20}). As a consequence,
for $d>4$, when one approaches $\beta_c$ from the strong coupling regime,  
one finds that $a_c=\sqrt{r_c}\neq 0$ and a first-order phase transition occurs. 
 The solution to Eq.~(\ref{s4e20}), on the critical point, subject to 
  Eq.~(\ref{s4e4}) and Eq.~(\ref{s4e5}), can be found numerically. In order to have 
a better behaved kernel when $d$ is close to $4$ we define a new function $H$,
\begin{equation}
H_c(z) = \frac{\rho_c(z)}{ z \sqrt{(b-z)(z-a)}}\;.
\end{equation}
With $\lambda = \frac{2 \pi }{d-2}$, the integral equation Eq.~(\ref{s4e20}),  
and the  constraints, Eqs.~(\ref{s4e4})-(\ref{s4e5}), and the equation which 
determines  $\beta_c$, Eq.~(\ref{s4e19}), become
\begin{equation}
\label{inteq}
\lambda H_c(z) =  \;\; 
 {\int}_a^b \;  \frac{dy}{ y+z } \;  
H_c(y) \;\frac{\sqrt{(b-y)(y-a)}}{\sqrt{(b+y)(y+a)}}\;,
\end{equation}
\begin{equation}
 {\int}_a^b \; dz \; H_c(z) \; z \sqrt{b-y)(y-a)} \; = \;1\;,
\end{equation}
\begin{equation}
{\int}_a^b \; dz \; H_c(z) \; \sqrt{b-y)(y-a)} \; = \;1\;,
\end{equation}
\begin{equation}
{1 \over \beta_c } = 2 ( d-2 ) {\int}_a^b \; dz  \;
 H_c(z) \; { \sqrt{(b-z)(z-a)} \over \sqrt{(b+z)(z+a)} } \; .
\end{equation}

The solution of the integral equation Eq.~(\ref{inteq}) can be
done numerically by  discretizing the kernel.
After the discretization, the problem is
reduced to an eigenvalue problem of a real non symmetric matrix. 
There are several ways to discretize the kernel. Any rule of
numerical integration is a discretization rule for the kernel.
It is known that for integral equations the best discretization rules
are the Gauss quadrature rules~\cite{Delves}.
There are several Gauss quadrature rules. We used the simplest 
possible: the Gauss-Chebyshev rule. All these rules
require to map the integration interval to $[-1,1]$.
Thus we perform the following change of variables 
\begin{equation}
 z = { b-a \over 2} \zeta + { a + b  \over 2 }\;,
\end{equation}
or
\begin{equation}
\zeta =  { 2 \over b-a} z - { a + b  \over b-a }\;.
\end{equation}
Under this change of variables the integral equation becomes
\begin{equation}
\lambda H_c(\zeta) =  \; {\int}_{-1}^{1} \; d{\xi} \; 
\frac{H_c(\xi)}{\xi + \zeta + 2 \frac{1+\kappa}{1-\kappa}} \; 
\frac{ \sqrt{1-\xi^2}}{\sqrt{(\xi + \frac{3+\kappa}{1-\kappa})
                             (\xi +\frac{1+3 \kappa}{1-\kappa})}} \; ,
\label{finteq}
\end{equation}
where $ \kappa= a/b$.
The constraints take the following form
\begin{equation}
{\int}_{-1}^{1} \; d{\xi} \; H_c(\xi) \; ( \xi + \frac{1+\kappa}{1-\kappa} )
\sqrt{1-\xi^2} = ({ 2 \over 1-\kappa })^3 { 1 \over b^3 }\;,
\label{fcon1}
\end{equation}
\begin{equation}
{\int}_{-1}^{1} \; d{\xi} \; H_c(\xi) \; \sqrt{1-\xi^2} = 
({ 2 \over 1-\kappa })^2  { 1 \over b^2 }\;,
\label{fcon2}
\end{equation}  
and the equation for $\beta_c$ is 
\begin{equation}
{1 \over \beta_c } = b ( d-2 )(1-\kappa) {\int}_{-1}^{1} \; d\xi  \;
 H_c(\xi) \; { \sqrt{1-\xi^2} \over
               \sqrt{(\xi + \frac{3+\kappa}{1-\kappa})
                     (\xi +\frac{1+3 \kappa}{1-\kappa})}  } \;.
\label{fbetac}
\end{equation}

The solution to Eq.~(\ref{finteq}) can be found up to an overall constant $C$,
(assuming that $\lambda$ is a nondegenerate eigenvalue). This
constant and the upper bound, $b$, of the support of $\rho_c$ can be
computed using the constrains  Eq.~(\ref{fcon1}), Eq.~(\ref{fcon2}).
Let's denote an eigenfunction of (\ref{finteq}) by $\hat{H_c}$. Then
$H_c$, which is the function that is positive in [-1,1] and satisfies the
constraints, (\ref{fcon1}) and (\ref{fcon2}),  is related to $\hat{H_c}$ by
$H_c = C \hat{H_c}$. If we now define
\begin{equation}
I_1= {\int}_{-1}^{1} \; d{\xi} \; \hat{H_c}(\xi) \; 
     ( \xi + \frac{1+\kappa}{1-\kappa})\sqrt{1-\xi^2} \;, 
\label{I1}
\end{equation}
\begin{equation}
I_2 = {\int}_{-1}^{1} \; d{\xi} \; \hat{H_c}(\xi) \; \sqrt{1-\xi^2}\;, 
\label{I2}
\end{equation} 
then 
\begin{equation}
b= { I_2 \over I_1} \, { 2 \over 1-\kappa}\;, 
\label{b}
\end{equation}
\begin{equation}
C = { I_1^2 \over I_2^3}\;.
\label{C}
\end{equation}

From the above formulas it is obvious that one can fix $\kappa$,
solve Eq.~(\ref{finteq}) and then find the eigenvalue $\lambda$ which
has a positive definite eigenfunction. Because the problem is
well defined one expects that there exists only one such function.
This expectation is confirmed by the numerical results.
It turns out that the eigenfunction with the largest eigenvalue
is the one which is positive definite in $[-1,1]$. The $n$th eigenfunction
has $n-1$ zeros in $[-1,1]$.
Thus for a given $\kappa$ one computes $b(\kappa)$ ,  $a(\kappa)$ ,
 $\lambda(\kappa)$ ,  $\beta_c(\kappa)$ , 
$d(\kappa) = 2 \pi \lambda(\kappa)^{-1} + 2$.

Using the above  numerical method we have computed $a_c,b_c$ with
great precision for $d$ in the interval (4.4,250).  Combining the numerical 
results with the analytical for $d\leq 4$,
$a_c$ and $b_c$ are plotted in Fig.~\ref{acbc} as functions of $1/d$ for
$d\geq 2$.  
The functions $a_c(d)$, $b_c(d)$ are continuous functions of $d$ at $d=4$.

Several interesting features now emerge from an analysis of this
data. On the one hand we can fit the functions $a_c(d)$, $b_c(d)$ with
great accuracy as power series of $d-4$ around $d=4$ and they agree
with the corresponding weak coupling expressions up to very high
orders.  On the other hand, if one does a careful extrapolation of
$a_c, b_c$ to $d=4$ a new feature is seen ( Figure.~\ref{logacbc}).
The upper limit, $b_c(d)$, extrapolates linearly in $d-4$ to $\pi$,
consistent with analyticity in the $d-4$ series expansion.  The data
alone determines the intercept $b_c(4)$ to be $\pi$ to an accuracy of
$10^{-8}$. When one examines the lower limit $a_c$, it also approaches
zero as $d$ approaches $4$. However it does not go to zero as a simple
power.  The more we improved our data near $d=4$ the higher the
effective power became.  It appears that $a_c$ may have an essential
singularity at $d=4$ vanishing faster than any power. Since the
discontinuity of the internal energy on the first order line is given
by $d a_c^2/(4(d-1))$, this is pertinent to the critical properties at
the end of the first order transition. The log-log plots of $a_c(d)$
and $\pi - b_c(d)$ in Fig.~\ref{logacbc} clearly support these
observations.

In  Fig.~\ref{acbc} we have also been helped by an expansion of
the functions ${(a_c+b_c)/ 2}$ and $a_c b_c$ in powers of $1/d$.
The coefficients of this expansion have been determined by best fits
on the numerical results and found to be consistent with integer
numbers within the precision of our determination.  This result is
also consistent with the results of the $1/d$ expansion which we shall
discuss in the next section. In particular we found
\begin{eqnarray}
\frac{a_c+b_c}{2} &=& 1 + \frac{2}{d} +
                      \frac{2}{d^2} -
                      \frac{6}{d^4} -
                      \frac{20}{d^5} -
                      \frac{48}{d^6} -
                      \frac{92}{d^7} -
                      \frac{118}{d^8}-
                      \frac{42}{d^9} +  .... \\
a_c b_c &=& 1-        \frac{12}{d} +
                      \frac{40}{d^2} -
                      \frac{8}{d^3} -
                      \frac{40}{d^4} -
                      \frac{128}{d^5} -
                      \frac{328}{d^6} -
                      \frac{694}{d^7} -
                      \frac{1112}{d^8} +  ....
\end{eqnarray}
The last integer term in both equations is uncertain.

Furthermore from the numerical computation of the $\beta_c$ for $d>4$, we can
see the $d<4$ result $\beta_c=1/d$ still holds above the critical point. We
performed the numerical calculations in double precision and we see no
deviation at all from the $1/d$ law. The deviation of $|\beta_c - 1/d|$ from
zero is determined to be less than $10^{-16}$ for $d$ in the interval
(4.4,250).

\section{Summary and Conclusion}

The 1/N expansion of matrix models has recently been used as a discrete
representation for summing over random surfaces and, through the
``double-scaling" limit, for studying low-dimensional string theories.  Even
more importantly, the large-$N$ expansion has provided us with a scheme for
addressing non-perturbative issues in non-Abelian gauge theories. For
instance, many qualitative features of QCD, {\it e.g.}, confinement, the OZI
rule, etc, can best be understood in a large-$N$ setting.

However, quantitative progress in these directions has been slow due
partly to the technical difficulties associated with the large number
of independent ``loop" variables in this limit. Nevertheless, it has
been possible to gain useful insights into various interesting
situations by utilizing as guides solvable models involving a small
number of matrices, {\it e.g.}, models involving two Hermitian
matrices.  Much less is known for models involving unitary
matrices. Our current work not only adds to the list of solvable
models in this category but also introduces new techniques for
addressing matrix model studies in the large-$N$ limit.

In this paper, we have studied the large-$N$ structure of simplicial
chiral models defined on a {$d-1$} dimensional simplex as one varies
$d$ and the coupling $\beta$. By exploring the global $U(N)\times
U(N)$ symmetry and by introducing an auxiliary complex matrix field,
we are able to reduce the problem to that of solving for the
eigenvalue distribution of a single Hermitian semi-positive-definite
matrix in the large-$N$ limit.  In addition to providing exact
large-$N$ solutions for several specific values of $d$, we are able to
identify and solve the strong-weak criticality problem for all values
of $d$, $0\leq d < \infty$.

For $0\leq d\leq 4$, analytic solutions for $\rho_c$ can be
found. Interestingly we find that the criticality occurs precisely at
$\beta_c=1/d$, as suggested by our previous
studies~\cite{RossiTan}. We find that the transition is third
order. For $4<d<\infty$, the criticality can also be studied by
solving a homogeneous integral equation.  However, we are only able to
carry this out numerically. Within numerical accuracy, we have shown
that the criticality again takes place at $\beta_c=1/d$, but with a
first order transition.

Since we are able  to reduce a $d-1$
dimensional  simplicial chiral model to a model involving a single complex
matrix, with $d$ entering as a parameter in the effective action, the
large-$N$ limit can thus be solved by finding a density function for
the eigenvalue distribution. Unlike usual matrix models where all
eigenvalues lie in a single connected band in the large-$N$ limit, this model
effectively involves two bands, a ``right-band" where
$\lambda_i>0$ and a ``left-band" where
$\lambda_j<0$. However, unlike other two-band problems~\cite{DDJT},
the distribution over these two bands are correlated.  This new feature presents
a challenge which cannot be handled by a conventional large-$N$ treatment.

 Our key result is the reduction of the above problem to that of
solving a single inhomogeneous integral equation for the eigenvalue
distribution of a single Hermitian semi-positive definite
matrix. Although we could not find a closed form solution to this
equation for arbitrary $d$, we are able to solve it in several
interesting special cases and we set up a systematic numerical
approach to the solutions. We have found that the critical surface is
defined by $\beta_c = 1/d$ for all $d$.

For small  $d$, $0<
\lowercase{d} < 4$, the models exhibit the third  order Gross Witten transition.
Indeed for $d = 1,2,3$ they coincide exactly with the chiral chains
studied earlier by Brower, Rossi and Tan~\cite{BRT}. In this region,
the criticality is related to that of $O(n)$ spin models on random
surfaces, as discussed by Gaudin-Kostov~\cite{Gaudin}. For
$\lowercase{d} > 4$, however, there is a first order transition ending
at the ``upper critical'' dimensions $\lowercase{d}= 4$. It therefore
appears that, from the perspective of the double-scaling limit, the
most interesting situation corresponds to $0<d\leq 4$. We have found
that the point $d=4$, having a logarithmic singularity, corresponds to
$\alpha=0$. In the language of the double-scaling limit, this
corresponds to having a vanishing ``string susceptibility",
$\alpha\equiv \gamma_{string} =0$, where $C_{sing}\sim
(\beta-\beta_c)^{-\gamma_{string}}$, which formally correspond to that
resulted from a $c=1$ CFT theory.  This calls for further studies
around $d=4$ which can provide further insight into possible different
mechanisms for generating $c=1$ physics. One way is to vary $d$ near
$d=4$. Another approach is to stay at $d=4$, and embellish the model
by relaxing the "permutation symmetry" of the original $d=4$
simplicial chiral model. This will be presented in a subsequent
publication.

\acknowledgments

We are deeply indebted with Prof. G. Cicuta for bringing
Ref.~\cite{Gaudin} to our attention and for useful conversations. This
work was supported in part by the U. S. Department of Energy, under
grant DE-FG02-91ER400688, Task A.

\appendix{Critical properties of chiral chains for $L=2,3,4,\infty$} 
\label{appa}

Chiral chain models are defined by the partition function 
\begin{equation}
Z_L\;=\;\int\prod_{i=1}^LdU_i \,\exp\left[
N\beta \sum_{i=1}^L {\rm Tr} (U_iU_{i+1}^\dagger +U_i^\dagger U_{i+1})
\right]\;,
\label{ae1}
\end{equation}
with $U_i=U_{i+L}$.
Free energy density, internal energy and specific heat are given by
\begin{eqnarray}
&&F_L\;=\; {1\over LN^2}\ln Z_L\;,\nonumber \\
&&U_L\;=\; {1\over 2}{\partial F_L\over \partial\beta}\;,\nonumber \\
&&C_L\;=\; \beta^2{\partial U_L\over \partial \beta}\;.
\label{ae1b}
\end{eqnarray}

When $L\rightarrow\infty$, $Z_L$ can be reduced to the partition function of
the Gross-Witten single-link problem~\cite{Gross}
\begin{equation}
Z\;=\;\int dU\,\exp \left[
N\beta\,{\rm Tr}(U+U^\dagger)\right]\;,
\label{ae2b}
\end{equation}
thus sharing the same thermodynamic properties.
The free energy density at $N=\infty$, $F={1\over N^2}\ln Z$, is
piecewise analytic with a third order transition at $\beta=\beta_c=1/2$
between the strong coupling and weak coupling domains.
The large-$N$ limit of the specific heat is
\begin{eqnarray}
C_\infty&=& \beta^2 \;\;\;\;\;\;\;\; {\rm for}\;\;\;\beta\leq \beta_c\;,\nonumber \\
C_\infty&=& {1\over 4} \;\;\;\;\;\;\;\; {\rm for}\;\;\;\beta\geq \beta_c.
\label{ae5}
\end{eqnarray}
The behavior of $C_\infty$ around $\beta_c$ can be characterized by a
specific heat critical exponent $\alpha=-1$.  It is worth noting that
an analysis of the double scaling limit, $N\rightarrow\infty$ and
$\beta\rightarrow\beta_c$, allows the determination of the correlation
length critical exponent, $\nu=3/2$~\cite{Periwal,DT}, and that
$\alpha$ and $\nu$ satisfy a hyperscaling relationship associated to a
two-dimensional critical phenomenon, $2\nu=2-\alpha$.  This fact is
related to the equivalence of the double scaling limit with the
continuum limit of a two-dimensional gravity model with central charge
$c=-2$.

It has been shown that in the context of single-matrix models the
parameter $N$ plays a role quite analogous to the volume in ordinary
systems, and double scaling limit turns out to be very similar to
finite size scaling in a two-dimensional critical
phenomenon~\cite{Carlson,Brezin}.  As a manifestation of this fact, it
has been observed that in the Gross-Witten single-link problem, (i)
the asymptotic approach of the complex $Z(N,\beta)$ zeroes closest to
$\beta_c$, $\bar{\beta}(N)$, toward the real axis occurs at a rate
determined by the correlation length exponent~\cite{Heller},
\begin{equation}
{\rm Im} \,\bar{\beta}(N)\;\propto\; N^{-{1\over \nu}}\;,
\label{ae6}
\end{equation}
and,  (ii) for sufficient large $N$ the position of the peak of the
specific heat, $\beta_{peak}(N)$, behaves as~\cite{largeN}
\begin{equation}
\beta_{peak}(N)-\beta_c\;\propto\;N^{-{1\over\nu}}\;.
\label{ae7}
\end{equation}
We recall that in ordinary critical behaviors finite size scaling
leads to relations of the type (\ref{ae6}-\ref{ae7}) with $N$
replaced by the size of the system.

The $L=2$ chiral chain again corresponds to a Gross-Witten model with
$\beta$ replaced by $2\beta$, thus obtaining $\beta_c=1/4$ and the
same critical exponents.

Solutions of the models with $L=3,4$ have been found  in Ref.~\cite{BRT}.
The results we present in the following without details on the derivations
are easily obtainable  from the analysis of Ref.~\cite{BRT}.

We recall that the chiral chain with $L=3$ is equivalent to the simplicial
model with $d=3$. In this case $\beta_c=1/3$ and the phase transition is 
still third order. In the weak coupling region, $\beta\geq\beta_c$,
the $N=\infty$ specific heat is given by
\begin{equation}
C_3^{({\rm w})}\;=\;\beta^2 + {1\over 8} - \beta^2\left( 1+{1\over 6\beta}\right)
\left( 1-{1\over 3\beta}\right)^{1/2}\;.
\label{ae8}
\end{equation}
Therefore close to $\beta_c$, 
\begin{equation}
C_3^{({\rm w})}\;=\; {17\over 72}-{1\over 2\sqrt{3}}
(\beta-\beta_c)^{1/2}+O(\beta-\beta_c)\;.
\label{ae9}
\end{equation}
Similarly in the strong coupling region, $\beta\leq\beta_c$, and close to
$\beta_c$,
\begin{equation}
C_3^{({\rm s})}\;=\; {17\over 72}-{1\over 2\sqrt{3}}
(\beta_c-\beta)^{1/2}+O(\beta_c-\beta)\;.
\label{ae10}
\end{equation}
Then the strong and weak coupling expressions of $C_3$ show that
the critical point $\beta_c=1/3$ is third
order and $\alpha=-1/2$.

For $L=4$ the study of the critical behavior around $\beta_c=\pi/8$ is slightly
subtler. 
In the weak coupling domain  the $N=\infty$ internal energy can be expressed as
\begin{equation}
U_4^{({\rm w})}\;=\;2\beta - {1\over 8\beta} - 2\beta\delta^2\;,
\label{ae11}
\end{equation}
where $\delta$ is implicitly determined  by the equation
\begin{equation}
{8\beta\over \pi}\left[ E\left(\sqrt{1-\delta^2}\right)-\delta^2 F\left(
\sqrt{1-\delta^2}\right)\right]\;=\;1\;.
\label{ae12}
\end{equation}
(This equation comes from the normalization of the eigenvalue
distribution $\rho(\theta)$ introduced in Ref.~\cite{BRT}).
Since $\delta=0$ at $\beta_c=\pi/8$, in order 
to study  the critical behavior close to $\beta_c$ 
we expand Eq.~(\ref{ae12}) around $\delta=0$, 
obtaining the following relation
\begin{equation}
{\beta-\beta_c\over \beta_c}\;=\; {1\over 2}\delta^2\left(
\ln {4\over \delta}+{1\over 2}\right) + O\left( \delta^4\right)\;,
\label{ae13}
\end{equation}
and therefore $\delta^2\sim\beta-\beta_c$ apart from logarithms.
Furthermore we have
\begin{equation}
{d\delta^2\over d\beta} \;=\; {16\over \pi \ln(4/\delta)}
\;+\;O(\delta^2)\;.
\label{ae14}
\end{equation}
We then obtain for the specific heat 
\begin{equation}
C_4^{({\rm w})}\;=\; {\pi^2\over 32}+{1\over 8}- {\pi^2\over 16\ln(4/\delta)}+
O\left(\beta-\beta_c\right)
\label{ae15}
\end{equation}
when $\beta\rightarrow\beta_c^+$.

A similar analysis can be performed in strong coupling,
where the internal energy can be written as
\begin{equation}
U_4^{({\rm s})}\;=\;2\beta - {1\over 8\beta} +2\beta {\zeta^2\over 1-\zeta^2}\;,
\label{ae16}
\end{equation}
with $\zeta$ implicitly defined by the equation
\begin{equation}
{8\beta\over \pi} {1\over \sqrt{1-\zeta^2}}
E\left( \sqrt{1-\zeta^2}\right)\;=\;1\;.
\label{ae17}
\end{equation}
Since $\zeta=0$ at $\beta_c$, we expand Eq.~(\ref{ae17}) around $\zeta=0$ obtaining
\begin{equation}
{\beta_c-\beta\over \beta_c}\;=\; {1\over 2}\zeta^2\left(
\ln {4\over \zeta}+{1\over 2}\right) + O\left( \zeta^4\right)\;.
\label{ae18}
\end{equation}
Consequently, $\zeta^2\sim\beta_c-\beta$ apart from logarithms and
\begin{equation}
{d\zeta^2\over d\beta} \;=\; -{16\over \pi \ln(4/\zeta)}
\;+\;O(\zeta^2)\;.
\label{ae19}
\end{equation}
We then obtain when $\beta\rightarrow\beta_c^-$, 
\begin{equation}
C_4^{({\rm s})}\;=\; {\pi^2\over 32}+{1\over 8}-{\pi^2\over 16\ln(4/\zeta)}+
O(\beta_c-\beta)\;.
\label{ae20}
\end{equation}

A comparison of Eqs.~(\ref{ae15}) and (\ref{ae20}) leads to 
the conclusion that the phase transition is again third order 
with a critical exponent $\alpha=0^-$.
The  critical exponent $\nu$ could then be determined
by using the 2-d hyperscaling relationship, obtaining $\nu=1$.
This value of $\nu$ was confirmed
by a numerical Monte Carlo study of
the scaling of the specific heat peak position at finite $N$,
we indeed observed a behavior like Eq.~(\ref{ae7}) compatible with $\nu=1$ within
a few per cent of uncertainty.

In Fig.~\ref{LC} we plot the specific heat versus $\beta$ for
$L=2,3,4,\infty$.  In conclusion we have seen that $L=2,3,4,\infty$
have a third order phase transition at increasing critical values
$\beta_c={1\over 4},{1\over 3},{\pi\over 8},{1\over 2}$, with specific
heat critical exponents $\alpha=-1,-1/2,0^-,-1$, respectively.  Notice
the behavior of $\alpha$ with respect to $L$, which is increasing for
$L=2,3,4$ reaching the limit of a third order critical behavior, but
then in large-$L$ limit it returns to $\alpha=-1$.

\appendix{Strong coupling expansion of chiral chain models}
\label{appb}

Strong coupling series of the free energy density of chiral chain models
are best generated by means of the character expansion, which 
leads to the following result
\begin{equation}
F_L(\beta)\;=\; F(\beta)\,+\, \widetilde{F}_L(\beta)\;,
\label{be1}
\end{equation}
where $F(\beta)$ is the free energy of the single unitary matrix model
($F(\beta)={1\over N^2}\ln Z$ and $Z$ is given by Eq.~(\ref{ae2b})),
and
\begin{equation}
\widetilde{F}_L\;=\;{1\over LN^2}\ln \sum_{(r)}d_{(r)}^2\,z_{(r)}(\beta)^L\;,
\label{be1b}
\end{equation}
$\sum_{(r)}$ denotes the sum over all irreducible representations of $U(N)$,
$d_{(r)}$ and $z_{(r)}(\beta)$ are the corresponding dimensions and character 
coefficients.
The calculation of the strong coupling 
series of $F_L(\beta)$ is much simplified in the large-$N$ limit, due to 
the following relationships~\cite{SC} 
\begin{equation}
F(\beta)\;=\;\beta^2\,+\,O\left(\beta^{2N+2}\right)\;,
\label{be3}
\end{equation}
and 
\begin{equation}
z_{(r)}(\beta)\;=\; \bar{z}_{(r)}\,\beta^n \,+\, O\left(\beta^{2N}\right)\;,
\label{be4}
\end{equation}
where $\bar{z}_{(r)}$ is independent of $\beta$ and
$n$ is the order of the representation $(r)$.
Explicit expressions of $d_{(r)}$ and $\bar{z}_{(r)}$ are given in
Ref.~\cite{SC}.
Notice that the large-$N$ strong coupling expansion of $\widetilde{F}_L(\beta)$ 
is actually a series in $\beta^L$,
\begin{equation}
\widetilde{F}_L\;=\;\sum_n c(n,L)\beta^{nL}\;.
\label{be5}
\end{equation}
$\widetilde{F}_L$ represents also the generating functional
for the ``potentials'' $W(n,L)$ introduced in Ref.~\cite{SC},
in the context of the strong coupling expansion of more general models,
indeed the following relationship holds
\begin{equation}
W(n,L)\;=\;{L\over 2} c(n,L)\;.
\label{be6}
\end{equation}
It is important to recall that the large-$N$ character coefficients have jumps
and singularities at $\beta={1\over 2}$~\cite{Green}, and therefore the relevant
region for a strong-coupling character expansion is $\beta<{1\over 2}$.

We have analyzed the strong coupling series of chiral chain models in
order to investigate their large-$N$ critical behaviors for
$L>4$. Given the simple behavior of the large-$N$ limit of $F(\beta)$,
we considered only the contributions from $\widetilde{F}(\beta)$, thus
working with series in $\beta^L$.  We generated about 15 terms for
each $L<10$ and analyzed, as series in $\beta^L$, the specific heat
derivative, which diverges at the critical point in a third phase
transition.  We employed the integral approximant
technique~\cite{Guttmann,Hunter,Fisher}, which at present seems to be
one of the most powerful method of resummation.  In particular we
considered integral approximants obtained from first order linear
differential equations.

Let us begin with the results obtained for the known cases $L=3,4$.
For $L=3$ already 15 terms in the series (in $\beta^L$) suffice to get
$\beta_c={1\over 3}$ and $\alpha=-{1\over 2}$  with a precision of
about $10^{-9}$ and $10^{-7}$ respectively. However
it is worth noticing that in the analysis of the specific heat derivative
we found spurious non-diverging singularities 
on the positive real axis for $\beta<\beta_c$.

Concerning the $L=4$ case, it is known that the integral approximant
resummation analysis cannot reproduce an $\alpha=0^-$ singularity
type~\cite{Hunter} and therefore it is not really suitable to this
case.  Anyway, we obtained a good determination of $\beta_c$, we found
${\pi\over 8}$ up to about $10^{-5}$, and a rather stable but wrong
exponent, $\alpha\simeq -0.18$, which should somehow simulate the
logarithmic corrections found in the Appendix~\ref{appa}, given that
they cannot be generated by the differential equation solution.  Again
we found spurious non-diverging singularities for $\beta<\beta_c$.

The strong-coupling analysis starts giving new information when $L>4$.
Due to the persistent presence of spurious singularities, guided by
the $L=3,4$ analysis, in all cases we considered the first diverging
singularity on the positive real axis as an estimate of the true
critical point.  For $L=5$ we obtained quite stable results:
$\beta_c\simeq 0.43756$ and $\alpha\simeq -0.17$. We should say that
the $L=4$ analysis suggests some caution in accepting this estimate of
$\alpha$, it could still be a masked $\alpha=0^-$.  The analysis of
$L=6$ series gave a rather stable estimate of the critical point
$\beta_c\simeq 0.4737$, but unstable exponents (although negative and
small).  Similar results were found for $L\geq 7$: $\simeq 0.504$ for
$L=7$, $\simeq 0.526$ for $L=8$, $\simeq 0.57$ for $L=10$.  Notice
that, unlike the $L\leq 6$ cases, these values cannot be considered as
an estimate of the critical point.  They are indeed larger than
${1\over 2}$, that is out of the region where a strong coupling
analysis can be predictive, and therefore something else must happen
before, breaking the validity of the strong coupling expansion.  An
example of this phenomenon comes from the Gross-Witten single-link
model (recovered when $L\rightarrow\infty$), where the strong coupling
expansion of the $N=\infty$ free energy leads to an analytical
function not having singularity at all, $F(\beta)=\beta^2$, thus
$\beta_c={1\over 2}$ cannot be determined from a strong coupling
analysis.

Of course we cannot consider this analysis satisfactory, but from it
we may hint at a possible scenario.  As for $L\leq4$, for $L=5,6$,
that is when the estimate of $\beta_c$ coming from the above strong
coupling analysis is smaller than ${1\over 2}$ and therefore
acceptable, the term $\widetilde{F}(\beta)$ in Eq.~(\ref{be1}) should
be the one relevant for the critical properties, determining the
critical points and giving $\alpha\neq -1$ (maybe $\alpha=0^-$ as in
the $L=4$ case).  For $L\geq 7$ the critical point may not be a
singular point in strong or weak coupling, but just the point where
weak coupling and strong coupling curves meet each other.  This would
cause a softer phase transition with $\alpha=-1$, as for the
Gross-Witten single-link problem.  We expect $\beta_c<{1\over 2}$ also
for $L\geq 7$.

This scenario would be consistent with the analysis of Green and
Samuel~\cite{Green2}, who studied the behavior of the link determinant
(i.e. $\langle {\rm det} U_iU_{i+1}^\dagger \rangle$) to determine the
critical points. The values of $\beta_c$ we found for $L=5,6$ are
consistent with their estimates.

\appendix{$\lowercase{d}=4$ model: 
weak and strong coupling expansion and series analysis.}
\label{appc}

We have briefly discussed in Section~\ref{sec6} the possibility of
performing weak and strong coupling expansion in the $d=4$ model
starting from an expansion in the powers of $k^2$. Here we want to
give more details on the concrete implementation of this program.

Let us first of all for convenience define a few auxiliary functions
of $k^2$, the labels s and w are to remind the strong and weak
coupling expansions, which we shall treat on the same footing.  In
terms of standard elliptic integrals we define
\begin{equation}
n_{\rm w}(k^2)\;\equiv\; {2\over \pi}\int_0^k
{\sqrt{1-\zeta^2}\over \sqrt{k^2-\zeta^2}}d\zeta
\;=\; {2\over \pi} E(k)\;,
\label{ce1a}
\end{equation}
\begin{equation}
n_{\rm s}(k^2)\;\equiv\; {2\over \pi}\int_0^k
{\sqrt{k^2-\zeta^2}\over \sqrt{1-\zeta^2}}d\zeta
\;=\; {2\over \pi} \left[ E(k)-(1-k^2)K(k)\right]\;,
\label{ce1b}
\end{equation}
\begin{equation}
d_{\rm w}(k^2)\;\equiv\;{2\over \pi}K(k)n_{\rm s}(k^2)\;,
\label{ce2a}
\end{equation}
\begin{equation}
d_{\rm s}(k^2)\;\equiv\;{2\over \pi}K(k)k^2 n_{\rm w}(k^2)\;.
\label{ce2b}
\end{equation}
let us introduce the function 
\begin{equation}
p(k^2)\;=\; \left( {2\over\pi}\right)^2\int_0^k\sqrt{k^2-\zeta^2}
\sqrt{1-\zeta^2}\,\Pi(\zeta^2,k)d\zeta\;,
\label{ce3}
\end{equation}
and notice that, because of the properties of elliptic integrals and
of Eq.~(\ref{s6e29}) we may obtain from Eqs.~(\ref{s6e26}-\ref{s6e27})
\begin{equation}
8\beta\;=\; {n(k^2)^2\over d(k^2)-p(k^2)}
\label{ce4}
\end{equation}
holding for the proper choice of indices both in the weak and in the
strong coupling phase.  In order to set up our expansion we therefore
need power series representation for the functions ${2\over \pi}E(k)$,
${2\over \pi}K(k)$ and $p(k^2)$.  The elliptic integrals of the first
and second kind have simple known expansions,
\begin{equation}
{2\over \pi}K(k)\;=\; 1+\sum_{n=1}^\infty c_{0n}\left(k^2\right)^n\;,
\label{ce5a}
\end{equation}
\begin{equation}
{2\over \pi}E(k)\;=\; 1-\sum_{n=1}^\infty {c_{0n}\over 2n-1}\left(k^2\right)^n\;,
\label{ce5b}
\end{equation}
where
\begin{equation}
c_{0n}\;=\; { (2n!)^2\over 2^{4n}(n!)^4}\;.
\label{ce6}
\end{equation}
The elliptic integral of the third kind, needed in the construction of
$p(k^2)$, admits the following expansion in the powers of the first
argument,
\begin{equation}
{2\over\pi} \Pi(\zeta^2,k)\;=\;\sum_{l=0}^\infty \zeta^{2l} I_l(k^2)\;,
\label{ce7}
\end{equation}
where
\begin{equation}
I_l(k^2)\;=\;{2\over\pi}
\int_0^1 {\xi^{2l}\over \sqrt{ (1-\xi^2)(1-k^2\xi^2)}}d\xi
\label{ce8}
\end{equation}
satisfy the recursion equation
\begin{equation}
(2l+1)k^2 I_{l+1}(k^2)-2l(1+k^2)I_l(k^2)+(2l-1)I_{l-1}(k^2)\;=\;0\;, 
\label{ce9}
\end{equation}
with boundary condition,
\begin{eqnarray}
I_0(k^2)&=&{2\over\pi}K(k)\;,\nonumber \\
I_1(k^2)&=&{2\over\pi k^2}\left[ K(k)-E(k)\right]\;.
\label{ce10}
\end{eqnarray}
In turn we may show that
\begin{equation}
J_l(k^2)\;\equiv\; (k^2)^{-(l+1)}{2\over \pi} \int_0^k
\sqrt{k^2-\zeta^2} \sqrt{1-\zeta^2} \,\zeta^{2l}d\zeta
\label{ce11}
\end{equation}
are related to $I_l$ by
\begin{equation}
J_l(k^2)\;=\; I_l(k^2)-(1+k^2)I_{l+1}(k^2)+k^2I_{l+2}(k^2)\;.
\label{ce12}
\end{equation}
As a consequence the function $p(k^2)$ is completely determined once
$I_l(k^2)$ are known, by the relationship
\begin{equation}
p(k^2)\;=\; \sum_{l=0}^\infty I_l(k^2)J_l(k^2) (k^2)^{l+1}\;.
\label{ce13}
\end{equation}
The recursion Eq.~(\ref{ce9}) may be solved if we expand the functions
$I_l(k^2)$ in a power series of $k^2$,
\begin{equation}
I_l(k^2)\;\equiv\;\sum_{n=0}^\infty c_{ln}(k^2)^n\;.
\label{ce14}
\end{equation}
It is tedious but straightforward to verify that
\begin{equation}
c_{ln}\;=\; {(2n+2l)!\,(2n)!\over \left[ (n+l)! \,n!\,2^{2n+l}\right]^2}
\label{ce15}
\end{equation}
satisfy the recursion. 

One may also for convenience define
\begin{equation}
J_l(k^2)\;\equiv\;\sum_{n=0}^\infty d_{ln}(k^2)^n\;,
\label{ce16}
\end{equation}
and finds from Eq.~(\ref{ce12}) that
\begin{equation}
d_{ln}\;=\;-{c_{ln}\over 2(2n-1)(n+l+1)}\;.
\label{ce17}
\end{equation}
Direct substitution of the series expansions thus obtained in
Eq.~(\ref{ce4}) allows to construct an expansion of $\beta$ (or
${1/\beta}$ respectively) in powers of $k^2$ which a simple symbolic
manipulation program can easily extend to extremely high orders.

In order to make our analysis complete we must manage to extend our
discussion to the evaluation of a physical observable.  By recalling
Eq.~(\ref{s4e14}), we notice that the expression for the internal
energy in the $d=4$ case is
\begin{equation}
U\;=\;{1\over 48\beta^2} \int_a^b dz'\,\rho(z')(z'^2-r)-{1\over 3}-
{1\over 12\beta}\;,
\label{ce18}
\end{equation}
which gives rise in weak coupling to
\begin{equation}
U_{\rm w}\;=\; {4\over 3n_{\rm w}(k^2)^2}\int_0^k
d\zeta\, \rho_{\rm w}(\zeta)(1-\zeta^2)-{1\over 3}-
{2\over 3}{d_{\rm w}(k^2)-p_{\rm w}(k^2)\over n_{\rm w}(k^2)^2}\;,
\label{ce19}
\end{equation}
while in strong coupling we obtain
\begin{equation}
U_{\rm s}\;=\;{4\over 3n_{\rm s}(k^2)^2}\int_0^k
d\zeta\, \rho_{\rm s}(\zeta)(k^2-\zeta^2)-{1\over 3}-
{2\over 3}{d_{\rm s}(k^2)-p_{\rm s}(k^2)\over n_{\rm s}(k^2)^2}\;.
\label{ce20}
\end{equation}
By considering the explicit form of the functions $\rho_{\rm w}(\zeta)$
and $\rho_{\rm s}(\zeta)$ which we obtain from Eqs.~(\ref{s6e26}-\ref{s6e27}),
it is straightforward to parameterize $U$ by
\begin{equation}
U\;=\;{4\over 3}{k^2 r(k^2)\over \left[d(k^2)-p(k^2)\right]n(k^2)^2}
-{1\over 3}-{2\over 3}{d(k^2)-p(k^2)\over n(k^2)^2}\;,
\label{ce21}
\end{equation}
where the functions $r(k^2)$ can in turn be reconstructed, by using
the results presented in this Appendix, in terms of the functions
$I_l$ and $J_l$.  The results for the two regimes are
\begin{equation}
r_{\rm w}(k^2)\;=\;
\sum_{l=1}^\infty J_l(k^2)\left[
I_{l-1}(k^2)-I_l(k^2)\right] (k^2)^l
\label{ce22}
\end{equation}
and
\begin{equation}
r_{\rm s}(k^2)\;=\;
\sum_{l=1}^\infty J_l(k^2)\left[
I_{l-1}(k^2)-k^2I_l(k^2)\right] (k^2)^l
\label{ce23}
\end{equation}
respectively.
By expanding Eq.~(\ref{ce21}) in power series of $k^2$, inverting Eq.~(\ref{ce4})
and substituting $k^2$ as a function of ${1/\beta}$ or $\beta$ respectively,
we obtain the standard weak and strong coupling series for the internal energy.

In conclusion we present some terms of the strong and weak coupling series
of the internal energy,
obtained by implementing the procedure outlined in this appendix:
\begin{eqnarray}
U_{\rm s}&&=\; 
\beta + 2 \,\beta^2  + 2 \,\beta^3  + 4 \,\beta^5  + 28 \,\beta^6  + 38\, \beta^7  
+ 8 \,\beta^8  + 440\, \beta^9  + 1936\, \beta^{10}+ 1712 \,\beta^{11}\nonumber \\
&&   + 4160 \,\beta^{12}   + 62160\, \beta^{13}
+ 178072 \,\beta^{14}   + 101038 \,\beta^{15}   +  
1215704 \,\beta^{16} + 9259720 \,\beta^{17}\nonumber \\
&&     + 17052880 \,\beta^{18}  
+ 15519376 \,\beta^{19}   +  291277184 \,\beta^{20}
+1351546592\,\beta^{21}+...
\label{ce24}
\end{eqnarray}

\begin{eqnarray}
U_{\rm w}&&=\;
1-{0.5\over 4\beta} -{0.03125\over(4\beta)^2}
-{0.012695\over (4\beta)^3}-{0.006744\over(4\beta)^4}
-{0.004131\over (4\beta)^5}-{0.002767\over (4\beta)^6}
\nonumber \\
&&-{0.001971\over (4\beta)^7}-
{0.001469\over (4\beta)^8}
-{0.001133\over (4\beta)^9}-
{0.000898\over (4\beta)^{10}}-
{0.000727\over (4\beta)^{11}}-
{0.000600\over (4\beta)^{12}}+...
\label{ce25}
\end{eqnarray}
The strong coupling series was also generated by using
the more general approach of Ref.~\cite{SC}, finding the same
results.

\appendix{Numerical Approach based on large $\lowercase{d}$ Expansion}
\label{appd}

The Ansatz (\ref{s8e10}) can also be used as a starting point for
numerical approximations based on the very simple consideration that
real analytic functions of $\zeta$ can be approximated with any
assigned precision by polynomials of sufficiently high degree in
$\zeta$ itself.  We may therefore introduce the $n$-th truncations of
$Q(\zeta)$ and $R(\zeta)$ respectively by the definitions
\begin{eqnarray}
Q_n(\zeta)\;=\;\sum_{i=1}^n q_i\zeta^i\;, \nonumber \\
R_n(\zeta)\;=\;\sum_{i=1}^n r_i\zeta^i\;.
\label{s9e1}
\end{eqnarray}
We may now explicitly perform a Laurent series expansion around the point
$\zeta=0$ in the form,
\begin{equation}
dQ_n(\zeta)\left[ 1-\sqrt{ 1-{c_1^{(n)}\over d\zeta}
-{c_2^{(n)}\over d\zeta^2}}\right]\;\equiv\;
\sum_{i=1}^n p_i\zeta^i + p_0 + \sum_{j=1}^\infty p_{-j}\zeta^{-j}\;,
\label{s9e2}
\end{equation}
where the coefficients $p_i$, $p_0$ and $p_{-j}$ are completely
determined in terms of $q_i$, $c_1$ and $c_2$. 

We may now notice that the asymptotic condition on $\phi_n(\zeta)$ forces us
to impose the relationships
\begin{eqnarray}
&&p_i+r_i\;=\;0\;\;\;\;\;\;\;i>0\;,\nonumber\\
&&p_0\;=\;0\;,\nonumber \\
&&p_{-1}\;=\;1\;.
\label{s9e3}
\end{eqnarray}
At this stage $R_n$,$c_1^{(n)}$,$c_2^{(n)}$ are completely determined
in terms of the coefficients $q_i$ and the parameter
$\tilde{z}^{(n)}$.  We may now consider the effect of substituting
$\phi_n(\zeta)$ into Eq.~(\ref{s8e9}) and power-series expanding in
$\zeta$; if we define $\phi_n^{(k)}(-2\tilde{z})$ to be the $k$-th
derivative of the function $\phi_n$ evaluated at the point
$-2\tilde{z}$, we may turn Eq.~(\ref{s8e9}) into the following
approximate relationship,
\begin{equation}
{\tilde{z}\over 2\beta}-d+{\zeta\over 2\beta}\;=\;
2\left[ R_n(\zeta)+dQ_n(\zeta)\right]+
(d-2)\sum_{k=0}^n {(-\zeta)^k\over k!}\phi_n^{(k)}(-2\tilde{z})\;,
\label{s9e4}
\end{equation}
which in turn decomposes into $n+1$ equations in the $n+1$ unknowns
$q_i,\tilde{z}$,
\begin{eqnarray}
&&{\tilde{z}\over 2\beta}-d\;=\;(d-2)\phi_n(-2\tilde{z})\;,
\nonumber\\
&&{1\over 2\beta}\;=\;2(dq_1-p_1)-(d-2)\phi_n^{(1)}(-2\tilde{z})\;,
\nonumber\\
&&0\;=\; 2(dq_k-p_k)+(d-2){(-1)^k\over k!} \phi_n^{(k)}(-2\tilde{z})
\;\;\;\;\;\;\;\;1< k\leq n\;.
\label{s9e5}
\end{eqnarray}
These equations may be solved numerically for arbitrary values of
$\beta$ and $d$ and offer better and better approximations to the true
eigenvalue distribution (by applying Eq.~(\ref{s8e19}) with increasing
values of $n$).  Numerical experiments have shown that when $\beta\gg
1/d$ even extremely small values of $n$ give quite accurate
predictions, while around the criticality, corresponding in this
language to the condition
\begin{equation}
\phi(-\tilde{z})\;=\;-1\;, 
\label{s9e6}
\end{equation}
which determines $\beta_c$, but the accuracy is definitely weaker.
This fact did not prevent us from determining the location of criticality,
by the use of $n\leq 8$, in the cases $d=3,4,5$, with a precision of about 1\%.

In the context of this discussion it is important to observe that
(even approximate) knowledge of $\phi(\zeta)$ implies an (approximate)
knowledge of the moments of the eigenvalue distribution, which may be
obtained from the Laurent series expansion via the relationship
\begin{equation}
f(z)\;=\;\sum_{n=0}^\infty {1\over z^{n+1}}\int_a^b
dz'\, z'^n \rho(z')\;=\;
\phi(z-\tilde{z})\simeq \sum_{j=1}^\infty
p_{-j}(z-\tilde{z})^{-j} \;,
\label{s9e7}
\end{equation}
which in particular implies that
\begin{equation}
\int_a^bdz'\,z'^2\rho(z')\;=\;\tilde{z}^2+2\tilde{z}p_{-2}+p_{-3}\;, 
\label{s9e8}
\end{equation}
and in turn we may extract the internal energy
via the relationship
\begin{equation}
d(d-1)U\;=\;{1\over 4\beta^2}\left(
\tilde{z}^2+2\tilde{z}p_{-2}+p_{-3}\right) -d -{1\over \beta}\;.
\label{s9e9}
\end{equation}

% ========================= References  =========================

\vfill
\pagebreak

\figure{ $d C$ versus $d\beta$ for the $d=2,3,4,\infty$ 
                    simplicial models. \label{dC} }

\figure{ $a_c$ and $b_c$  versus $1/d$. The crosses mark the 
                       points corresponding to $d=4$. \label{acbc} } 

\figure{ Log-log plot of  $a_c$ and $\pi-b_c$  versus $d-4$.\label{logacbc} }

\figure{ Specific heat versus $\beta$ for the $L=2,3,4,\infty$
                      chiral chain models. \label{LC} } 

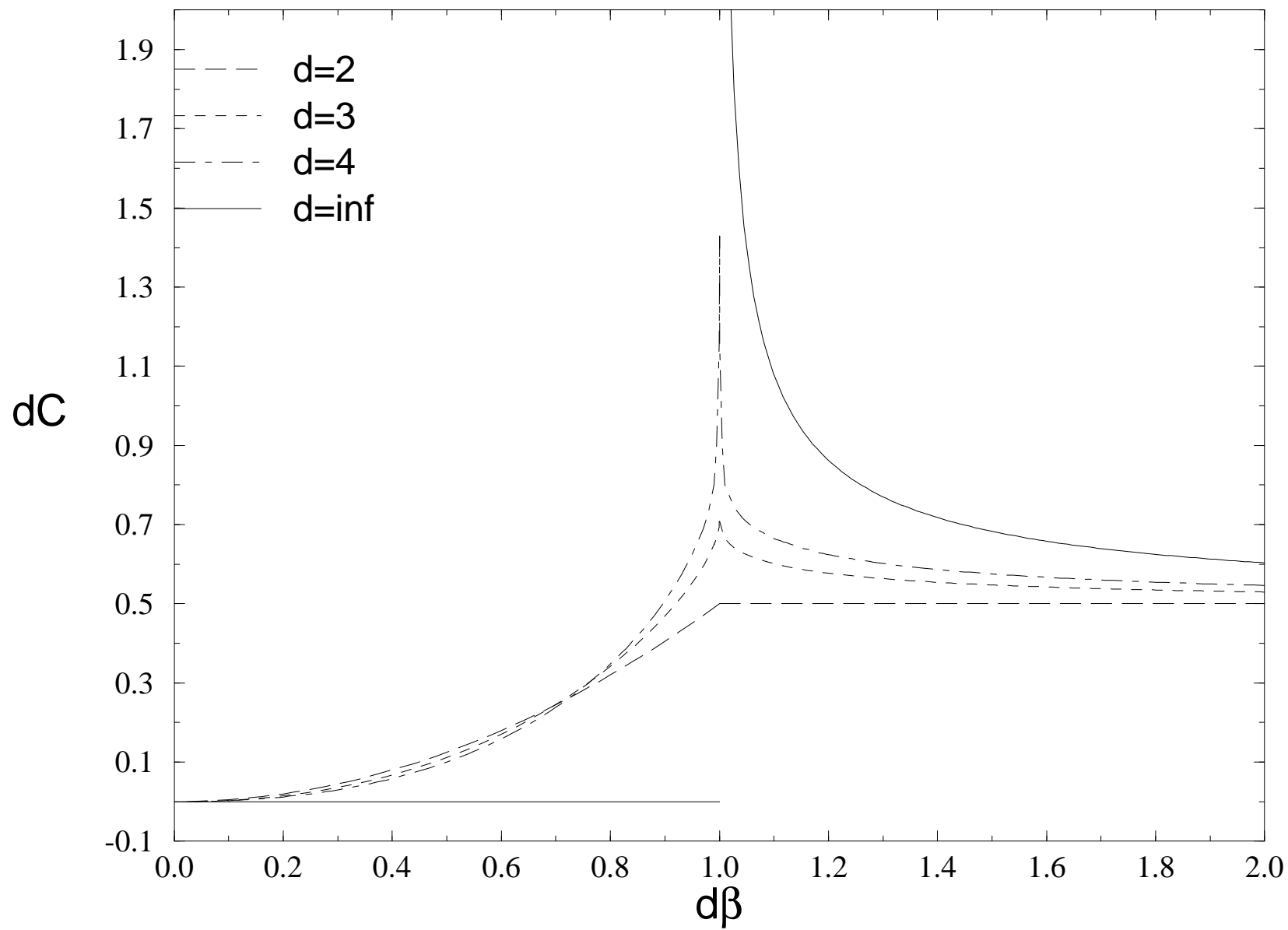
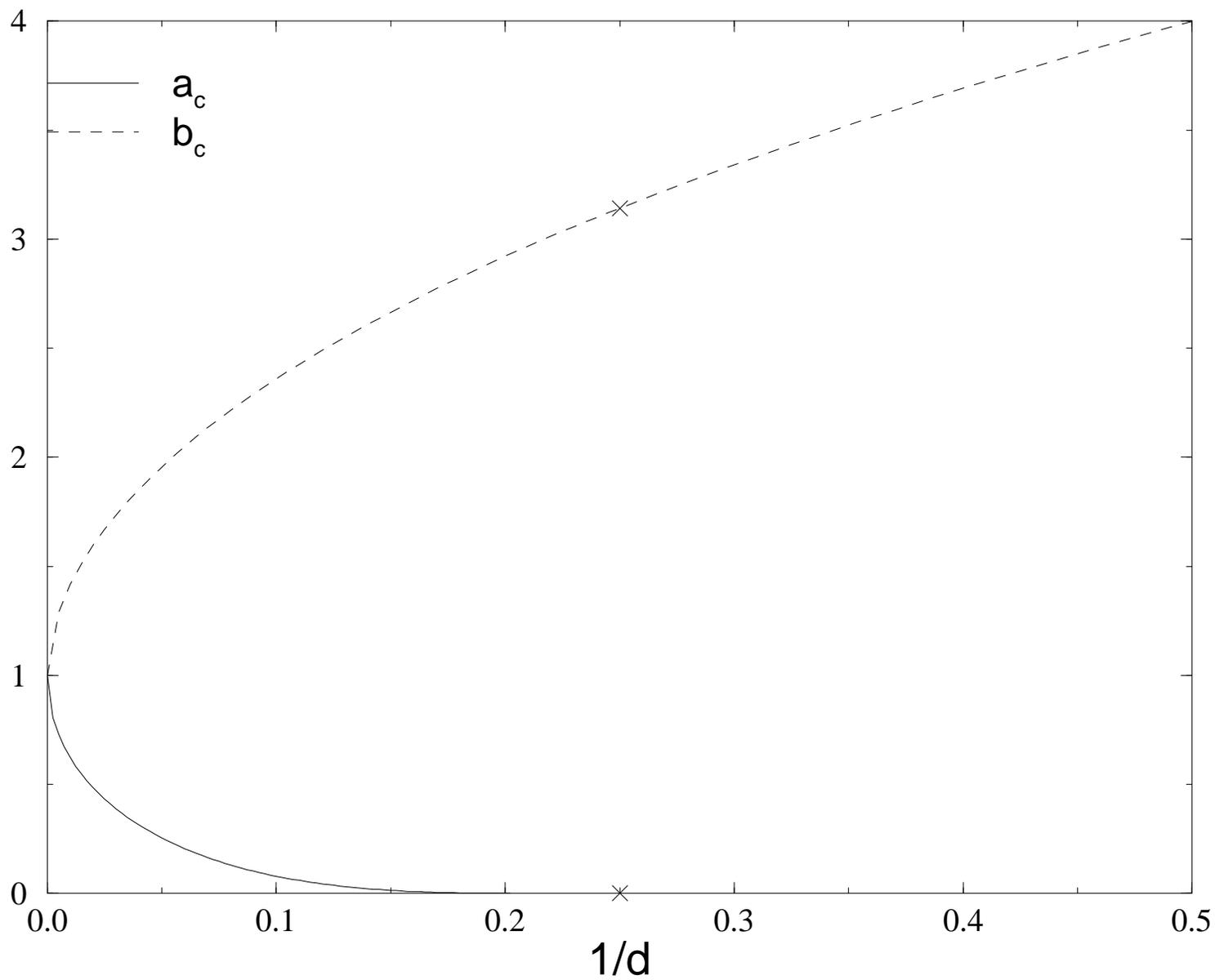
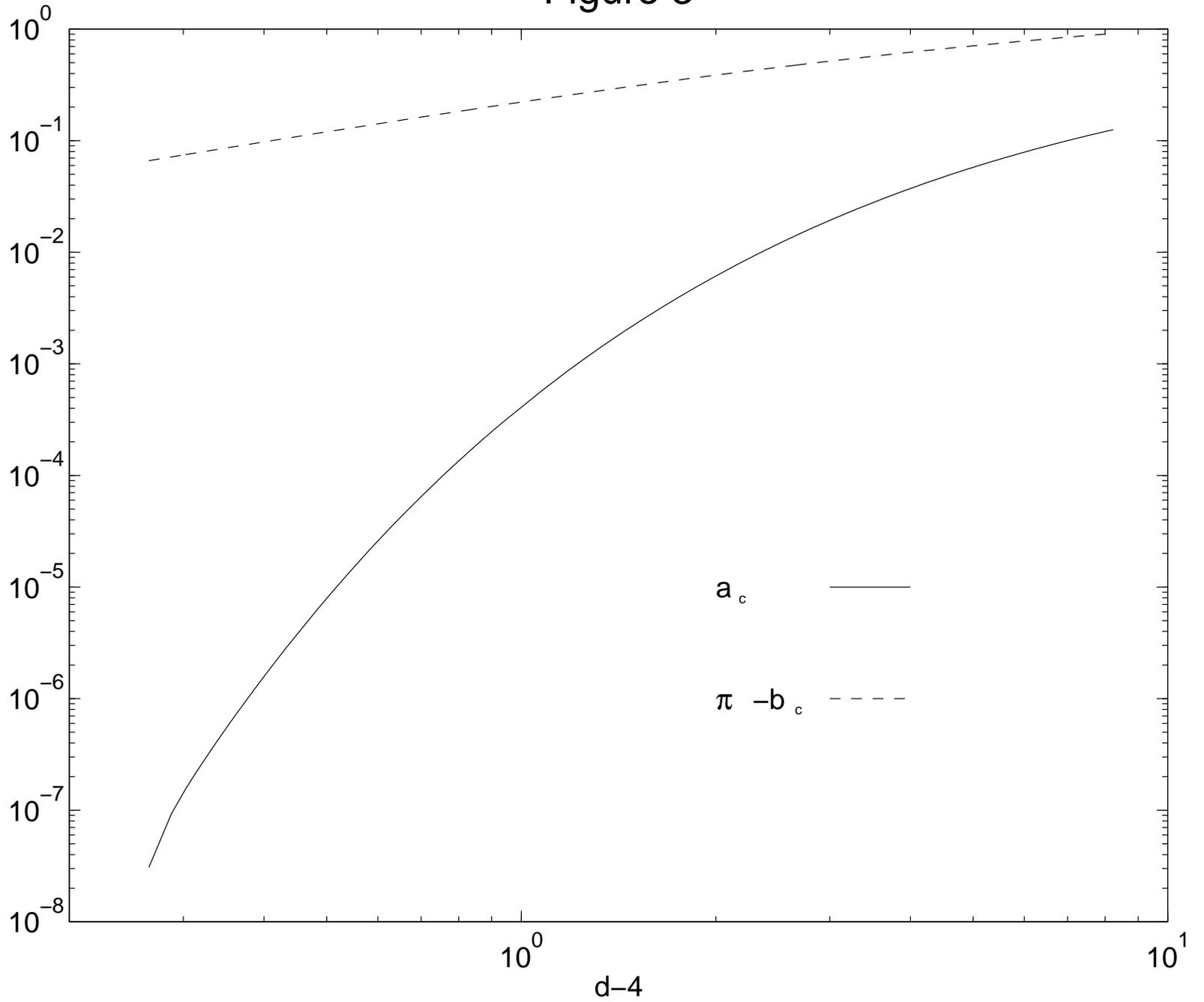
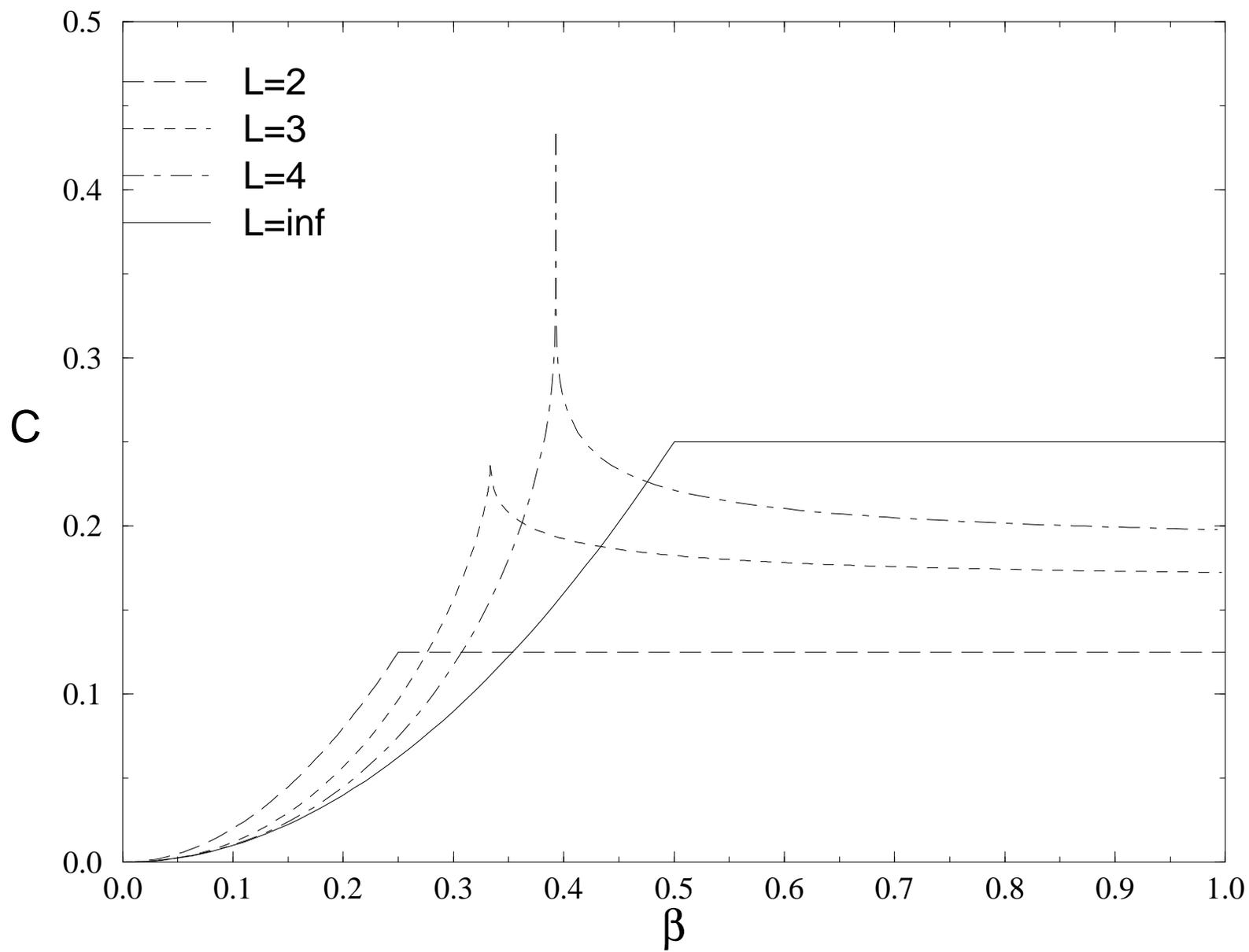
\end{document}